\documentclass[aps, prb, reprint, showpacs, twocolumn, 10pt, groupedaddress]{revtex4-1}

\usepackage{graphicx, amsmath, dsfont, dcolumn, bm, times, color, braket, ulem}

\begin{document}

\bibliographystyle{apsrev4-1}

\title{Phonon-mediated decay of singlet-triplet qubits in double quantum dots}

\author{Viktoriia Kornich}
\affiliation{Department of Physics, University of Basel, Klingelbergstrasse 82, CH-4056 Basel, Switzerland}
\author{Christoph Kloeffel}
\affiliation{Department of Physics, University of Basel, Klingelbergstrasse 82, CH-4056 Basel, Switzerland}
\author{Daniel Loss}
\affiliation{Department of Physics, University of Basel, Klingelbergstrasse 82, CH-4056 Basel, Switzerland}

\date{\today}

\begin{abstract}
We study theoretically the phonon-induced relaxation ($T_1$) and decoherence times ($T_2$) of singlet-triplet qubits in lateral GaAs double quantum dots (DQDs). When the DQD is biased, Pauli exclusion enables strong dephasing via two-phonon processes. This mechanism requires neither hyperfine nor spin-orbit interaction and yields $T_2 \ll T_1$, in contrast to previous calculations of phonon-limited lifetimes. When the DQD is unbiased, we find $T_2 \simeq 2 T_1$ and much longer lifetimes than in the biased DQD. For typical setups, the decoherence and relaxation rates due to one-phonon processes are proportional to the temperature $T$, whereas the rates due to two-phonon processes reveal a transition from $T^2$ to higher powers as $T$ is decreased. Remarkably, both $T_1$ and $T_2$ exhibit a maximum when the external magnetic field is applied along a certain axis within the plane of the two-dimensional electron gas. We compare our results with recent experiments and analyze the dependence of $T_1$ and $T_2$ on system properties such as the detuning, the spin-orbit parameters, the hyperfine coupling, and the orientation of the DQD and the applied magnetic field with respect to the main crystallographic axes.
\end{abstract}


\maketitle

\let\oldvec\vec
\renewcommand{\vec}[1]{\ensuremath{\boldsymbol{#1}}}

\section{Introduction}
\label{sec:Introduction}

The spin states of quantum dots (QDs) are promising platforms for quantum computation.\cite{loss:pra98, kloeffel:annurev13} In particular, remarkable progress has been made with $S$-$T_0$ qubits in lateral GaAs double quantum dots (DQDs),\cite{petta:sci05, foletti:nphys09, shulman:sci12, levy:prl02, klinovaja:prb12} where a qubit is based on the spin singlet ($S$) and triplet ($T_0$) state of two electrons in the DQD. In this encoding scheme, rotations around the $z$ axis of the Bloch sphere can be performed on a subnanosecond timescale \cite{petta:sci05} through the exchange interaction, and rotations around the $x$ axis are enabled by magnetic field gradients across the QDs.\cite{foletti:nphys09}

The lifetimes of $S$-$T_0$ qubits have been studied with great efforts. When the qubit state precesses around the $x$ axis, dephasing mainly results from Overhauser field fluctuations, leading to short dephasing times $T_2^{*} \sim 10\mbox{ ns}$.\cite{khaetskii:prl02, merkulov:prb02, coish:prb04, petta:sci05, johnson:nat05, bluhm:prl10} This low-frequency noise can be dynamically decoupled with echo pulses,\cite{petta:sci05, barthel:prl10, bluhm:nphys11, medford:prl12} and long decoherence times $T_2 > 200\mbox{ $\mu$s}$ have already been measured.\cite{bluhm:nphys11} In contrast to $x$-rotations, precessions around the $z$ axis dephase predominantly due to charge noise.\cite{dial:prl13, higginbotham:prl14} Rather surprisingly, however, recent Hahn echo experiments by Dial \textit{et al.} \cite{dial:prl13} revealed a relatively short $T_2 \simeq \mbox{0.1--1 $\mu$s}$ and a power-law dependence of $T_2$ on the temperature $T$. The origin of the observed decoherence is 
so far unknown, although the dependence on $T$ suggests that lattice vibrations (phonons) may play an important role.   

In this work, we calculate the phonon-induced lifetimes of a $S$-$T_0$ qubit in a lateral GaAs DQD. Taking into account the spin-orbit interaction (SOI) and the hyperfine coupling, we show that one- and two-phonon processes can become the dominant decay channels in these systems and may lead to qubit lifetimes on the order of microseconds only. While the decoherence and relaxation rates due to one-phonon processes scale with $T$ for the parameter range considered here, the rates due to two-phonon processes scale with $T^2$ at rather high temperatures and obey power laws with higher powers of $T$ as the temperature decreases. Among other things, the qubit lifetimes depend strongly on the applied magnetic field, the interdot distance, and the detuning between the QDs. Based on the developed theory, we discuss how the lifetimes can be significantly prolonged. 

The paper is organized as follows. In Sec.~\ref{sec:SystemAndBasisStates} we present the Hamiltonian and the basis states of our model. In the main part, Sec.~\ref{sec:LargeDetuning}, we discuss the calculation of the lifetimes in a biased DQD and investigate the results in detail. In particular, we show that two-phonon processes lead to short dephasing times and identify the magnetic field direction at which the lifetimes peak. The results for unbiased DQDs are discussed in Sec.~\ref{sec:SmallDetuning}, followed by our conclusions in Sec.~\ref{sec:Conclusions}. Details and further information are appended.

\section{System, Hamiltonian, and Basis States}
\label{sec:SystemAndBasisStates}

We consider a lateral GaAs DQD within the two-dimensional electron gas (2DEG) of an AlGaAs/GaAs heterostructure that is grown along the [001] direction, referred to as the $z$ axis. Confinement in the $x$-$y$-plane is generated by electric gates on the sample surface, and the magnetic field $\bm{B}$ is applied in-plane to avoid orbital effects. When the DQD is occupied by two electrons, the Hamiltonian of the system reads 
\begin{eqnarray}
H &=& \sum_{j=1,2} \Bigl( H_0^{(j)} + H_{Z}^{(j)} + H_{\rm SOI}^{(j)} + H_{\rm hyp}^{(j)} + H_{\rm el-ph}^{(j)} \Bigr) \nonumber \\
& & + H_{C} + H_{\rm ph} , 
\end{eqnarray}
where the index $j$ labels the electrons, $H_0$ comprises the kinetic and potential energy of an electron in the DQD potential, $H_Z$ is the Zeeman coupling, $H_{\rm SOI}$ is the SOI, $H_{\rm hyp}$ is the hyperfine coupling to the nuclear spins, $H_{\rm el-ph}$ is the electron-phonon coupling, $H_C$ is the Coulomb repulsion, and $H_{\rm ph}$ describes the phonon bath. 

\begin{figure}[tb]
\begin{center}
\includegraphics[width=0.85\linewidth]{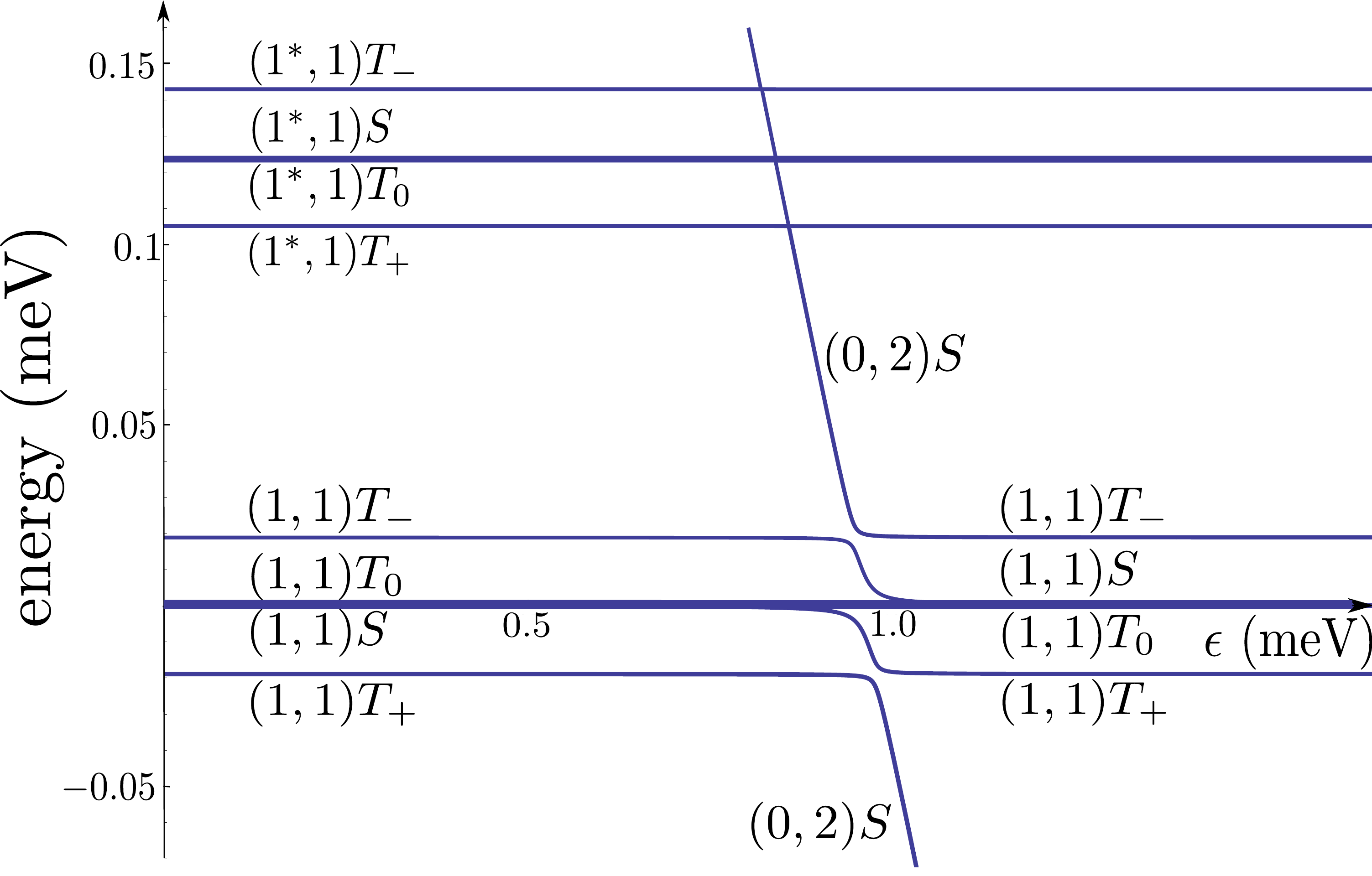}
\caption{The energy spectrum of the DQD calculated for the parameters described in the text. The $S$-$T_0$ qubit is formed by the eigenstates of type $|(1,1)S\rangle$ and $|(1,1)T_0\rangle$. }
\label{fig:spectrum}
\end{center}
\end{figure} 

The electron-phonon interaction has the form 
\begin{equation}
H_{\rm el-ph} = \sum_{\vec{q},s} W_s(\vec{q}) a_{\vec{q}s} e^{i\vec{q}\cdot\vec{r}} + \mbox{h.c.} ,
\label{eq:HelphBasicFormMainText}
\end{equation}
where $\vec{r}$ is the position of the electron, $\vec{q}$ is a phonon wave vector within the first Brillouin zone, $s \in \{l, t_1, t_2\}$ stands for the longitudinal ($l$) and the two transverse ($t_1, t_2$) phonon modes, and ``h.c.'' is the hermitian conjugate. The coefficient $W_s(\vec{q})$ depends strongly on $\bm{q}$ and $s$, and is determined by material properties such as the relative permittivity $\epsilon_r$, the density $\rho$, the speed $v_l$ ($v_t$) of a longitudinal (transverse) sound wave, and the constants $\Xi$ and $h_{14}$ for the deformation potential and piezoelectric coupling, respectively. The annihilation operator for a phonon of wave vector $\bm{q}$ and mode $s$ is denoted by $a_{\vec{q}s}$. The Hamiltonian 
\begin{equation}
H_{\rm SOI} = \alpha \left( p_{x'} \sigma_{y'} - p_{y'} \sigma_{x'} \right) + \beta \left( p_{y'} \sigma_{y'} - p_{x'} \sigma_{x'} \right)
\end{equation} 
contains both Rashba and Dresselhaus SOI. Here $p_{x'}$ and $p_{y'}$ are the momentum operators for the $x'$ and $y'$ axes, respectively. The latter coincide with the crystallographic axes [100] and [010], respectively, and $\sigma_{x'}$ and $\sigma_{y'}$ are the corresponding Pauli operators for the electron spin. We take into account the coupling to states of higher energy by performing a Schrieffer-Wolff transformation that removes $H_{\rm SOI}$ in lowest order.\cite{khaetskii:prb00, aleiner:prl01, golovach:prl04, golovach:prb08, stano:prb05, stano:prl06, raith:prl12} The resulting Hamiltonian $\widetilde{H}$ is equivalent to $H$, except that $H_{\rm SOI}$ is replaced by 
\begin{equation}
\widetilde{H}_{\rm SOI} \simeq g \mu_B (\vec{r}_{\rm SOI} \times \vec{B}) \cdot \vec{\sigma},
\end{equation}
where $g$ is the in-plane $g$ factor, $\bm{\sigma}$ is the vector of Pauli matrices, and 
\begin{equation}
\vec{r}_{\rm SOI} = \left( \frac{y'}{l_R} + \frac{x'}{l_D} \right) \vec{e}_{[100]} - \left( \frac{x'}{l_R} + \frac{y'}{l_D} \right) \vec{e}_{[010]}.
\end{equation} 
Here $x'$ and $y'$ are the coordinates of the electron along the main crystallographic axes, whose orientation is provided by the unit vectors $\vec{e}_{[100]}$ and $\vec{e}_{[010]}$, respectively. The spin-orbit lengths are defined as $l_R = \hbar/(m_{\rm eff} \alpha)$ and $l_D = \hbar/(m_{\rm eff} \beta)$, where $m_{\rm eff}$ is the effective electron mass in GaAs and $\alpha$ ($\beta$) is the Rashba (Dresselhaus) coefficient. For our analysis, the most relevant effect of the nuclear spins is the generation of an effective magnetic field gradient between the QDs, which is accounted for by $H_{\rm hyp}$. We note that this magnetic field gradient may also result from a nearby positioned micromagnet.\cite{laird:prl07, pioroladriere:nphys08, brunner:prl11} For details of $H$ and $\widetilde{H}$, see Appendix~\ref{sec:Hamiltonian}.  

The $S$-$T_0$ qubit in this work is formed by the basis states $\ket{(1,1)S}$ and $\ket{(1,1)T_0}$, where the notation $(m,n)$ means that $m$ ($n$) electrons occupy the left (right) QD. In first approximation, these states read \begin{eqnarray}
\ket{(1,1)S} &=& \ket{\Psi_{+}} \ket{S} , \\
\ket{(1,1)T_0} &=& \ket{\Psi_{-}} \ket{T_0} ,
\end{eqnarray} 
with 
\begin{equation}
\ket{\Psi_{\pm}} =  \frac{\ket{\Phi_L^{(1)} \Phi_R^{(2)}} \pm \ket{\Phi_R^{(1)} \Phi_L^{(2)}}}{\sqrt{2}} ,
\end{equation}
where the $\Phi_{L,R}(\bm{r})$ are orthonormalized single-electron wave functions for the left and right QD, respectively (see also Appendix~\ref{sec:BasisStates}).\cite{burkard:prb99, stepanenko:prb12} The spin singlet is 
\begin{equation}
\ket{S} = \frac{\ket{\uparrow\downarrow} -\ket{\downarrow\uparrow}}{\sqrt{2}} ,
\end{equation}
whereas 
\begin{equation}
\ket{T_0} = \frac{\ket{\uparrow\downarrow} + \ket{\downarrow\uparrow}}{\sqrt{2}} ,
\end{equation}
with the quantization axis of the spins along $\bm{B}$. Analogously, one can define the states $\ket{(1,1)T_+} = \ket{\Psi_{-}} \ket{\uparrow \uparrow}$ and $\ket{(1,1)T_-} = \ket{\Psi_{-}}\ket{\downarrow \downarrow}$, which are energetically split from the qubit by $\pm g \mu_B |\bm{B}|$. For our analysis of the phonon-induced lifetimes, a simple projection of $\widetilde{H}$ onto this 4D subspace of lowest energy is not sufficient, because 
\begin{equation}
\sum_j \Bigl( \bra{\Psi_+}H_{\rm el-ph}^{(j)} \ket{\Psi_+} - \bra{\Psi_-}H_{\rm el-ph}^{(j)} \ket{\Psi_-} \Bigr) = 0 . 
\end{equation} 
That is, corrections from higher states must be taken into account in order to obtain finite lifetimes.\cite{golovach:prb08, meunier:prl07} The spectrum that results from the states considered in our model is plotted in Fig.\ \ref{fig:spectrum}. Depending on the detuning $\epsilon$ between the QDs, the lifetimes of the qubit are determined by admixtures from $\ket{(2,0)S}$, $\ket{(0,2)S}$, or states with excited orbital parts.

\section{Regime of Large Detuning}
\label{sec:LargeDetuning}

\subsection{Effective Hamiltonian and Bloch-Redfield theory}
\label{secsub:EffHamAndBlochRedfield}

We first consider the case of a large, positive detuning $\epsilon$ at which the energy gap between $\ket{(0,2)S}$ and the qubit states is smaller than the orbital level spacing $\hbar \omega_0$. In this regime, contributions from states with excited orbital parts are negligible, and projection of $\widetilde{H}$ onto the basis states $|(1,1)T_0\rangle$, $|(1,1)S\rangle$, $|(1,1)T_+\rangle$, $|(1,1)T_-\rangle$, $|(0,2)S\rangle$, and $|(2,0)S\rangle$ yields  
\begin{widetext}
\begin{equation}
\widetilde{H} =
\begin{pmatrix}
    P_T & \frac{\delta b_B}{2} & 0 & 0 & 0 & 0  \\
    \frac{\delta b_B}{2} & V_+-V_-+P_T & \frac{\Omega}{\sqrt{2}} & -\frac{\Omega}{\sqrt{2}} & -\sqrt{2}t+P_{S}^\dagger & -\sqrt{2}t+P_{S} \\ 
    0 & \frac{\Omega}{\sqrt{2}} & E_Z + P_T & 0 & 0 & 0 \\
    0 & -\frac{\Omega}{\sqrt{2}} & 0 & -E_Z+P_T & 0 & 0 \\
   0 & -\sqrt{2}t+P_{S} & 0 & 0 & -\epsilon+U-V_-+P_{SR}& 0\\
   0 & -\sqrt{2}t+P_{S}^\dagger &0 & 0 & 0 & \epsilon+U-V_-+P_{SL} \\
  \end{pmatrix} + H_{\rm ph} .
\label{eq:matrix}
\end{equation}
\end{widetext}
Here $P_T$, $P_S$, $P_S^\dagger$, $P_{SL}$, and $P_{SR}$ are the matrix elements of the electron-phonon interaction, $t$ is the tunnel coupling,  $U$ is the on-site repulsion, $V_{\pm} = \langle \Psi_{\pm} | H_C | \Psi_{\pm}\rangle$, $E_Z = g \mu_B |\bm{B}|$, 
\begin{eqnarray}
\Omega = g \mu_B &\bigl(& \langle\Phi_L|(\vec{r}_{\rm SOI}\times\vec{B})_z|\Phi_L\rangle  \nonumber \\
& & - \langle\Phi_R|(\vec{r}_{\rm SOI}\times\vec{B})_z|\Phi_R\rangle \bigr) ,
\end{eqnarray}
and $\delta b_B=2\langle(1,1)S|H_{\rm hyp}|(1,1)T_0\rangle$ (see also Appendix \ref{secsub:HyperfineInteraction}). We note that the energy in Eq.\ (\ref{eq:matrix}) was globally shifted by $\bra{(1,1)T_0} \bigl( H_0^{(1)}+ H_0^{(2)} + H_C \bigr) \ket{(1,1)T_0}$. Furthermore, we mention that the state $\ket{(2,0)S}$ is very well decoupled when $\epsilon$ is large and positive. In Eq.~(\ref{eq:matrix}), $\ket{(2,0)S}$ is mainly included for illustration purposes, allowing also for large and negative $\epsilon$ and for an estimate of the exchange energy at $\epsilon \simeq 0$.

In order to decouple the qubit subspace $\{|(1,1)S\rangle, |(1,1)T_0\rangle\}$, we first apply a unitary transformation to $\widetilde{H}$ that diagonalizes $\widetilde{H} - \sum_{j} H_{\rm el-ph}^{(j)}$ exactly. Then we perform a third-order Schrieffer-Wolff transformation that provides corrections up to the third power in the electron-phonon coupling, which is sufficient for the analysis of one- and two-phonon processes. The resulting effective Hamiltonian can be written as $H_{\rm q} + H_{\rm q-ph}(\tau) + H_{\rm ph}$ in the interaction representation, where the time is denoted by $\tau$ to avoid confusion with the tunnel coupling. Introducing the effective magnetic fields $\vec{B_{\rm eff}}$ and $\vec{\delta B}(\tau)$ and defining $\vec{\sigma^\prime}$ as the vector of Pauli matrices for the $S$-$T_0$ qubit, 
\begin{equation}
H_{\rm q}= \frac{1}{2} g \mu_B \vec{B_{\rm eff}} \cdot \vec{\sigma^\prime}
\end{equation} 
describes the qubit and 
\begin{equation}
H_{\rm q-ph}(\tau) = \frac{1}{2} g \mu_B \vec{\delta B}(\tau)\cdot \vec{\sigma^\prime}
\end{equation} 
describes the interaction between the qubit and the phonons. The time dependence results from 
\begin{equation}
H_{\rm q-ph}(\tau) = e^{i H_{\rm ph} \tau/\hbar} H_{\rm q-ph} e^{-i H_{\rm ph} \tau/\hbar} .
\end{equation} 
For convenience, we define the basis of $\vec{\sigma^\prime}$ such that $B_{\textrm{eff},x} = 0 = B_{\textrm{eff},y}$. Following Refs.\ \onlinecite{golovach:prl04, borhani:prb06}, the decoherence time ($T_2$), the relaxation time ($T_1$), and the dephasing contribution ($T_{\varphi}$) to $T_2$ of the qubit can then be calculated via the Bloch-Redfield theory (see also Appendix~\ref{sec:BlochRedfieldTheory}), which yields 
\begin{gather}
\frac{1}{T_2} = \frac{1}{2 T_1} + \frac{1}{T_\varphi}, \\ 
\frac{1}{T_1} = J_{xx}^+(\omega_Z)+J_{yy}^+(\omega_Z) , \\ 
\frac{1}{T_{\varphi}} = J_{zz}^+(0) ,
\end{gather} 
where $\hbar \omega_Z = J_{\rm tot} = |g\mu_B \vec{B_{\rm eff}}|$ and 
\begin{equation}
J_{ii}^+(\omega)=\frac{g^2\mu_B^2}{2 \hbar^2}\int_{-\infty}^{\infty}\cos(\omega \tau)\langle\delta B_i(0)\delta B_i (\tau)\rangle d\tau .
\end{equation}
The correlator $\langle \delta B_i(0)\delta B_i (\tau) \rangle$ is evaluated for a phonon bath in thermal equilibrium and depends strongly on the temperature $T$.

\subsection{Input parameters}
\label{secsub:InputParameters}

The material properties of GaAs are $g = -0.4$, $m_{\rm eff} = 6.1 \times 10^{-32} \mbox{ kg}$, $\epsilon_r \simeq 13$, $\rho = 5.32 \mbox{ g/cm$^3$}$, $v_l \simeq 5.1\times 10^3 \mbox{ m/s}$ and $v_t \simeq 3.0\times 10^3 \mbox{ m/s}$ (see also Appendix \ref{secsubsub:DisplacementOperator}),\cite{cleland:book, adachi:properties, ioffe:data} $h_{14} \simeq -0.16 \mbox{ As/m$^2$}$,\cite{adachi:properties, huebner:pss73, ioffe:data} and $\Xi\approx -8 \mbox{ eV}$.\cite{adachi:gaas, vandewalle:prb89} In agreement with $\omega_0 /(2 \pi) = 30\mbox{ GHz}$,\cite{dial:prl13} we set $l_c = \sqrt{\hbar/(m_{\rm eff} \omega_0)} \simeq 96 \mbox{ nm}$, which is the confinement length of the QDs due to harmonic confining potential in the $x$-$y$ plane. For all basis states, the orbital part along the $z$ axis is described by a Fang-Howard wave function \cite{fang:prl66} of width $3 a_z = \mbox{6 nm}$ (see Appendix \ref{sec:BasisStates}). Unless stated otherwise, we set $l_R=2\ \mu\rm m$ and $l_D=1\ \mu\rm m$,\cite{hanson:rmp07, khaetskii:prb01, winkler:book} where $l_D$ is consistent with the assumed $a_z$ (see also Appendix \ref{sec:InputParameters}).\cite{winkler:book} We note, however, that adapting $a_z$ to $l_D$ is not required, because changing the width of the 2DEG by several nanometers turns out not to affect our results. All calculations are done for $|\vec{B}| = 0.7 \mbox{ T}$,\cite{bluhm:prl10, shulman:sci12} $\delta b_B = -0.14 \mbox{ $\mu$eV}$, in good agreement with, e.g., Refs.\ \onlinecite{bluhm:prl10, dial:prl13}, and an interdot distance of $2a = 400 \mbox{ nm}$. For Figs.\ \ref{fig:spectrum}--\ref{fig:angular_dependence} (large $\epsilon$), we use $U = 1 \mbox{ meV}$, $t = 7.25\mbox{ $\mu$eV}$, and $V_+ = 40\mbox{ $\mu$eV}$.\cite{stepanenko:prb12} We choose here $V_- = 39.78\mbox{ $\mu$eV}$ such that the resulting energy splitting $J_{\rm tot}(\epsilon)$ between the qubit states is mostly determined by the hyperfine coupling at $\epsilon \to 0$, as commonly realized experimentally.\cite{petta:sci05, dial:prl13} The detuning $\epsilon$ is then set such that $0 < U- V_\pm -\epsilon < \hbar \omega_0$ and $J_{\rm tot} = 1.43 \mbox{ $\mu$eV}$, and we note that this splitting is within the range studied in Ref.\ \onlinecite{dial:prl13}.

\subsection{Temperature dependence}
\label{secsub:TemperatureDependence}

\begin{figure}[tb] 
\begin{center}
\includegraphics[width=0.85\linewidth]{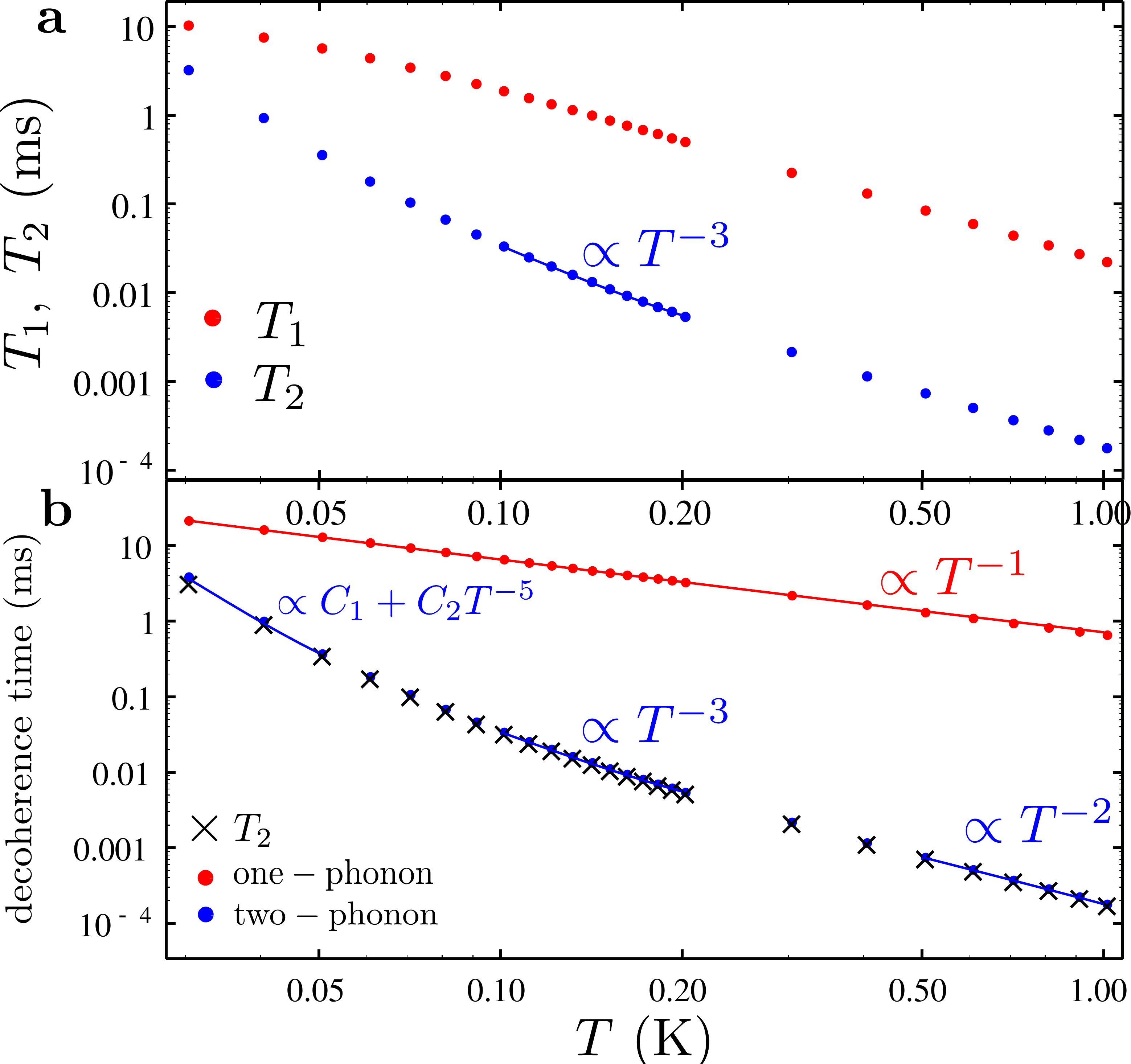}
\caption{(a) Temperature dependence of the decoherence time ($T_2$, blue) and relaxation time ($T_1$, red) for the parameters in the text. The solid line corresponds to a power-law fit to $T_2$ for $\mbox{0.1 K} \leq T \leq \mbox{0.2 K}$, which yields $T_2 \propto T^{-3}$ and good agreement with recent experiments.\cite{dial:prl13} We note that $T_2 \ll T_1$. (b) The decoherence time due to one-phonon ($1/\Gamma_2^{\rm 1p}$) and two-phonon processes ($1/\Gamma_2^{\rm 2p}$) and the full decoherence time $T_2 = 1/\Gamma_2= 1/(\Gamma_2^{\rm 1p} + \Gamma_2^{\rm 2p})$ as a function of temperature. We note that $1/\Gamma_2^{\rm 2p}$ changes its behaviour from $\propto C_1+C_2 T^{-5}$ to $\propto T^{-2}$, where $C_1$ and $C_2$ are constants, whereas $1/\Gamma_2^{\rm 1p} \propto T^{-1}$ for the range of $T$ considered here. }
\label{fig:tem_dep}
\end{center}
\end{figure} 

\begin{figure}[tb] 
\begin{center}
\includegraphics[width=0.85\linewidth]{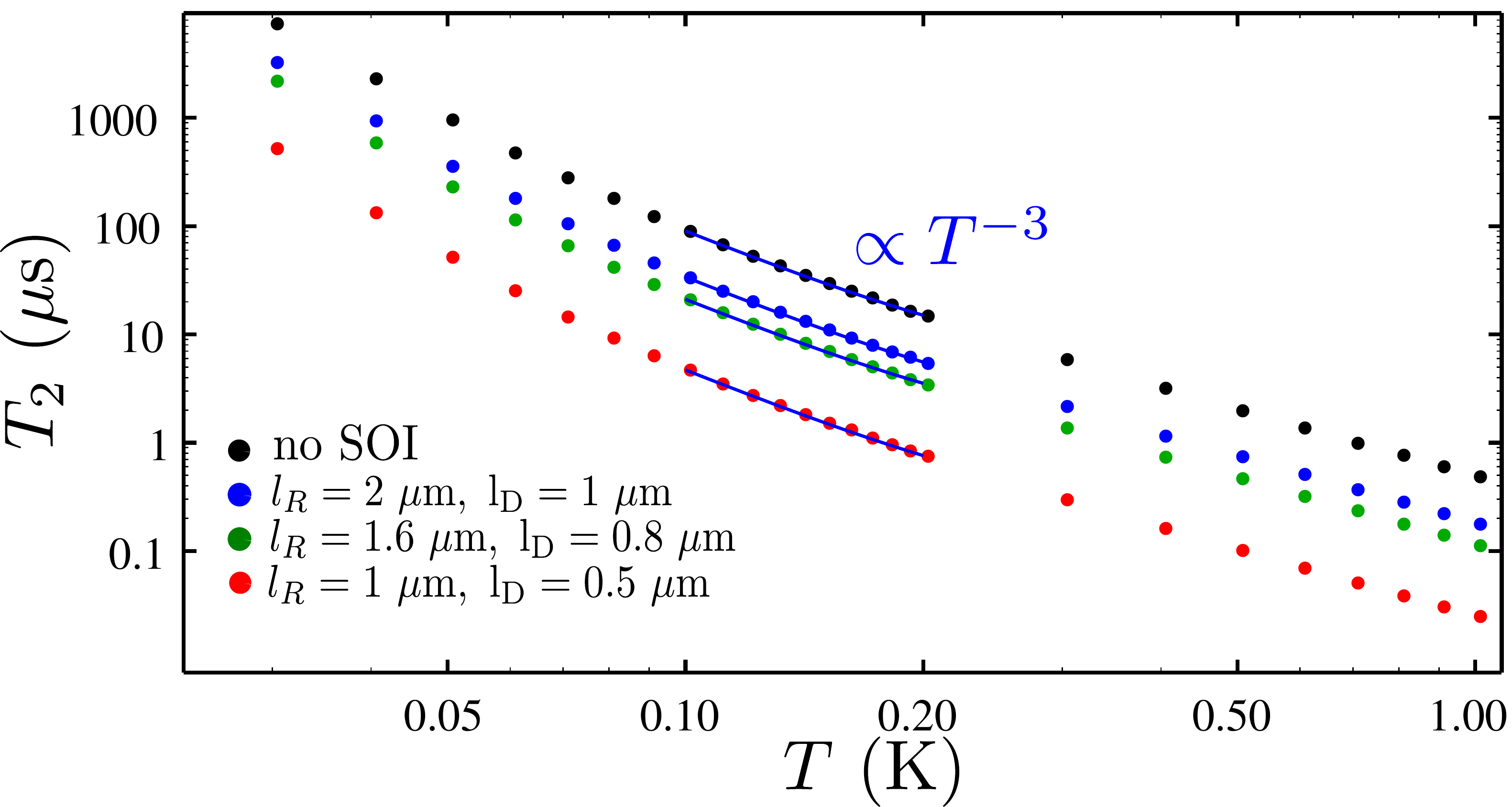}
\caption{Dependence of the decoherence time $T_2$ on the temperature for the parameters in the text and different spin-orbit lengths. Keeping the splitting $J_{\rm tot}$ between the qubit states constant, the values chosen for the detuning $\epsilon$ are \mbox{0.896 meV} (black), \mbox{0.912 meV} (blue), \mbox{0.918 meV} (green), and \mbox{0.933 meV} (red), increasing with increasing SOI. Within the range $T = \mbox{100--200 mK}$, $T_2 \propto T^{-3}$ in all cases. We note that the best quantitative agreement with the experiment \cite{dial:prl13} is obtained for the strongest SOI (red), where $l_R=1\ \mu\rm m$ and $l_D=0.5\ \mu\rm m$.}
\label{fig:different_spin_orbit}
\end{center}
\end{figure}

Figures \ref{fig:spectrum}--\ref{fig:different_spin_orbit} consider $\vec{B}$ applied along the $x$ axis that connects the two QDs, assuming that the $x$ axis coincides with the crystallographic $[110]$ direction. The geometry $x \parallel [110]$ is realized in most experiments,\cite{barthel:prl10, medford:prl12, higginbotham:prl14} particularly because GaAs cleaves nicely along [110]. In stark contrast to previous theoretical studies of phonon-limited lifetimes, where $T_2 = 2 T_1$,\cite{golovach:prl04, bulaev:prl05, trif:prl09, maier:prb13, hachiya:arX} Fig.\ \ref{fig:tem_dep}(a) reveals $T_2 \ll T_1$ at $30\mbox{ mK} \leq T \leq 1\mbox{ K}$ considered here, which implies $T_\varphi \ll T_1 $. In the discussion below we therefore focus on the details of the temperature dependence of $\Gamma_2 = 1/T_2$. We note, however, that the contributions to $\Gamma_2$ and $\Gamma_1 = 1/T_1$ from one-phonon processes scale similarly with $T$, and analogously for two-phonon processes. Defining $\Gamma_2^{\rm 1p}$ ($\Gamma_2^{\rm 2p}$) as the decoherence rate due to one-phonon (two-phonon) processes, Fig.\ \ref{fig:tem_dep}(b) illustrates $\Gamma_2^{\rm 2p} \gg \Gamma_2^{\rm 1p}$, and so $\Gamma_2 = \Gamma_2^{\rm 1p} + \Gamma_2^{\rm 2p} \simeq \Gamma_2^{\rm 2p}$. In the considered range of temperatures, we find $\Gamma_2^{\rm 1p} \propto T$. This behavior results from the fact that $\hbar \omega_{Z}/(k_B T)<1$ for our parameters, where $k_B$ is the Boltzmann constant. Therefore, the dominant terms in the formula for $\Gamma_2^{\rm 1p}$ are proportional to Bose-Einstein distributions defined as 
\begin{equation}
n_B(\omega) = \frac{1}{e^{\hbar \omega/(k_B T)} - 1 } 
\label{eq:BoseEinsteinDistribution}
\end{equation}
and may all be expanded according to $n_B(\omega) \simeq k_B T/(\hbar \omega)$, keeping in mind that the $n_B(\omega)$ contributing to $\Gamma_2^{\rm 1p}$ are evaluated at $\omega = \omega_Z$ because of energy conservation. The time $1/\Gamma_2^{\rm 2p}$ due to two-phonon processes smoothly changes its behaviour from $C_1+C_2 T^{-5}$ at $T \sim 40\mbox{ mK}$ to $T^{-2}$ with increasing temperature, where $C_n$ are constants. This transition is explained by the fact that, in the continuum limit, the rate corresponds to an integral over the phonon wave vector $\vec{q}$, where the convergence of this integral is guaranteed by the combination of the Bose-Einstein distribution and the Gaussian suppression that results from averaging over the electron wave functions. More precisely, the decay rate is obtained by integrating over the wave vectors of the two involved phonons. Due to conservation of the total energy, however, considering only one wave vector $\vec{q}$ is sufficient for this qualitative discussion. For $\Gamma_2^{\rm 2p}$, we find that the dominating terms decay with $\bm{q}$ due to factors of type 
\begin{equation}
f_s(\bm{q}) = e^{-(q_x^2+q_y^2)l_c^2} \hspace{0.02cm} n_B(\omega_{\bm{q}s}) \left[n_B(\omega_{\bm{q}s}) + 1 \right] ,
\end{equation}
where $q_x$ and $q_y$ are the projections of $\bm{q}$ onto the $x$ and $y$ axis, respectively, and $\hbar \omega_{\bm{q}s} = \hbar v_s |\bm{q}|$ is the phonon energy. Whether the Bose-Einstein part or the Gaussian part from $f_s(\bm{q})$ provides the convergence of the integral depends on $l_c$, $v_s \in \{ v_l, v_t \}$, and mainly $T$, as the latter can be changed significantly. When the Gaussian part $\exp[-(q_x^2+q_y^2)l_c^2]$ cuts the integral, $\Gamma_2^{\rm 2p} \propto T^2$ due to the expansion $n_B(n_B+1)\simeq (k_B T)^2/(\hbar \omega_{\bm{q}s})^2$ that applies in this case. When $n_B(n_B+1)$ affects the convergence of the integral, terms with higher powers of $T$ occur. The resulting temperature dependence is rather complex, but is usually well described by $1/\Gamma_2^{\rm 2p} = C_m + C_n T^{-\nu}$ with $\nu \geq 2$ for different ranges of $T$ [see Fig.\ \ref{fig:tem_dep}(b)]. The temperature ranges for the different regimes are determined by the details of the setup and the sample. For the parameters 
considered here, a power-law approximation $T_2 \propto T^{\eta}$ for $T = \mbox{100--200 mK}$ yields $\eta \simeq -3$ mainly because of the dephasing due to two-phonon processes (see Figs.\ \ref{fig:tem_dep} and \ref{fig:different_spin_orbit}), which agrees well with the experimental data of Ref.~\onlinecite{dial:prl13}. 

Figure \ref{fig:different_spin_orbit} shows the resulting temperature dependence of $T_2$ for different spin-orbit lengths. Remarkably, the calculation yields short $T_2$ even when SOI is completely absent. Keeping $J_{\rm tot} = 1.43\mbox{ $\mu$eV}$ fixed by adapting the value of $\epsilon$, one finds that $T_2$ decreases further with increasing SOI. As seen in Eq.~(\ref{eq:matrix}), $\widetilde{H}_{\rm SOI}$ couples $\ket{(1,1)S}$ to the triplet states $\ket{(1,1)T_+}$ and $\ket{(1,1)T_-}$. An important consequence of the resulting admixtures is that greater detunings are required in order to realize a desired $J_{\rm tot}$. In Fig.~\ref{fig:different_spin_orbit}, for instance, $\epsilon$ increases from $0.896\mbox{ meV}$ (no SOI) to $0.933\mbox{ meV}$ ($l_R = 1\mbox{ $\mu$m}$, $l_D = 0.5\mbox{ $\mu$m}$). As explained below, increasing $\epsilon$ decreases the lifetimes because it enhances the effects of $\ket{(0,2)S}$ through reduction of the energy gap (see also Fig.\ \ref{fig:spectrum}). 

\subsection{Origin of strong dephasing}
\label{secsub:OriginStrongDephasing}

The results discussed thus far have revealed two special features of the phonon-mediated lifetimes of $S$-$T_0$ qubits in biased DQDs. First, $T_2 \ll T_1$, as seen in Fig.~\ref{fig:tem_dep}(a). Second, the strong decay does not require SOI, as seen in Fig.~\ref{fig:different_spin_orbit}. These features have not been observed in previous calculations for, e.g., spin qubits formed by single-electron\cite{khaetskii:prb01, golovach:prl04} or single-hole\cite{bulaev:prl05, trif:prl09} or two-electron\cite{golovach:prb08} states in GaAs QDs, hole-spin qubits in Ge/Si nanowire QDs,\cite{maier:prb13} or electron-spin qubits in graphene QDs.\cite{hachiya:arX} Therefore, we discuss the dominant decay mechanism for $S$-$T_0$ qubits in DQDs in further detail and provide an intuitive explanation for our results. 

\begin{figure}[tb] 
\begin{center}
\includegraphics[width=0.85\linewidth]{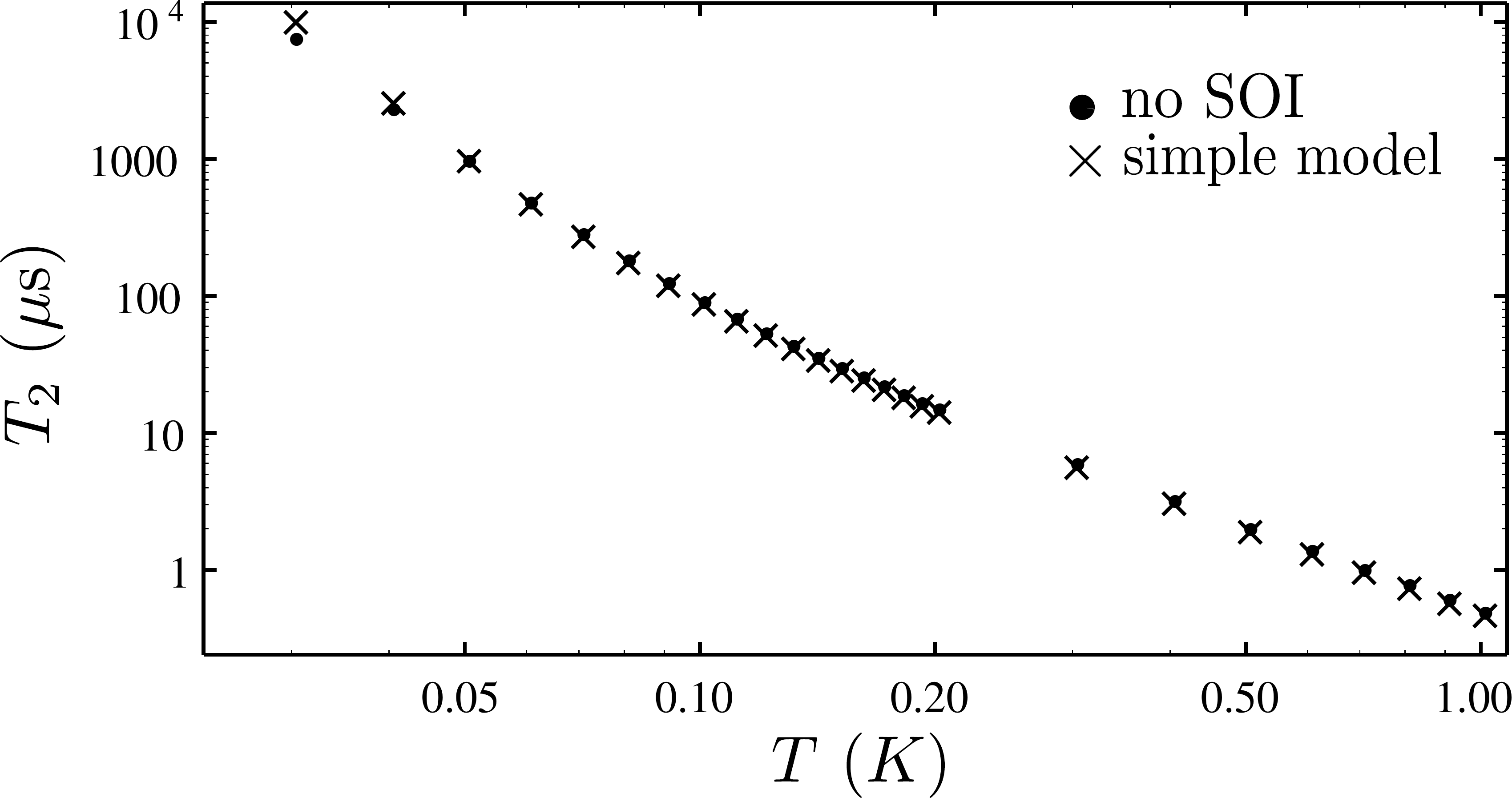}
\caption{Decoherence time $T_2$ as a function of temperature from two different models. The dotted line is also shown in Fig.~\ref{fig:different_spin_orbit} and was calculated via Eq.~(\ref{eq:matrix}), using the parameters in the text with $\Omega = 0$ (no SOI) and $\epsilon = 0.896\mbox{ meV}$. The crosses result from Eq.~(\ref{eq:Simple3x3DecRateFinalMainText}), using exactly the same parameters. We note that the associated $J_{\rm tot}$ differ only slightly. The remarkable agreement demonstrates that the simple model of Sec.~\ref{secsub:OriginStrongDephasing} accounts for the dominant decay mechanism. At $T \lesssim 50\mbox{ mK}$, the curves start to deviate because relaxation is no longer negligible. When the hyperfine coupling in Eq.~(\ref{eq:Simple3x3MatrixMainText}) is not omitted, excellent agreement is obtained also at low temperatures. }
\label{fig:ComparisonSimpleModel}
\end{center}
\end{figure}

Assuming again a large, positive detuning $\epsilon$, with $0 < U - V_{\pm} - \epsilon < \hbar \omega_0$, and setting $\Omega = 0$ (no SOI), the states $|(1,1)T_+\rangle$, $|(1,1)T_-\rangle$, and $|(2,0)S\rangle$ of Eq.\ (\ref{eq:matrix}) are practically decoupled from the qubit. The relevant dynamics are then very well described by
\begin{equation}
\widetilde{H} =
\begin{pmatrix}
    0 & \frac{\delta b_B}{2} & 0   \\
    \frac{\delta b_B}{2} & V_+-V_- & -\sqrt{2}t+P_{S}^\dagger  \\ 
   0 & -\sqrt{2}t+P_{S}  & -\epsilon+U-V_- + \widetilde{P}\\
  \end{pmatrix} + H_{\rm ph} ,
\label{eq:Simple3x3MatrixMainText}
\end{equation}
with $\ket{(1,1)T_0}$, $\ket{(1,1)S}$, and $\ket{(0,2)S}$ as the basis states and 
\begin{equation}
\widetilde{P} = P_{SR} - P_T .
\end{equation}
In the absence of SOI, the hyperfine interaction ($\delta b_B$) is the only mechanism that couples the spin states and enables relaxation of the $S$-$T_0$ qubit. We note that even when $\Omega$ is nonzero the relaxation times $T_1$ are largely determined by the hyperfine coupling instead of the SOI for the parameters considered in this work. At sufficiently large temperatures, where $T_2 \ll T_1$, $\delta b_B$ is negligible in the calculation of $T_2$, leading to pure dephasing, $T_2 = T_\varphi$. In addition, the matrix element $P_{S}$ turns out to be negligible for our parameters. Following Appendix~\ref{sec:SimpleModelDephasing}, we finally obtain
\begin{equation}
\frac{1}{T_2} = \frac{1}{T_\varphi} = \frac{2 t^4}{\hbar^2 (\Delta_S^\prime)^6} \int_{-\infty}^{\infty}  \langle\widetilde{P}^2(0)\widetilde{P}^2(\tau)\rangle d\tau 
\label{eq:Simple3x3DecRateFinalMainText}
\end{equation}    
from this simple model, where 
\begin{equation}
\Delta_S^\prime = \sqrt{(U - V_{+} - \epsilon)^2 + 8 t^2}
\end{equation}
corresponds to the energy difference between the eigenstates of type $\ket{(1,1)S}$ and $\ket{(0,2)S}$ (using $\delta_B = 0$). We note that terms of type $a_{\bm{q}s}^\dagger a_{\bm{q}s}$ and $a_{\bm{q}s} a_{\bm{q}s}^\dagger$ must be removed from $\widetilde{P}^2$ in Eq.~(\ref{eq:Simple3x3DecRateFinalMainText}), as the Bloch-Redfield theory requires $\langle \bm{\delta B}(\tau) \rangle$ to vanish (see also Appendix~\ref{sec:SimpleModelDephasing}).\cite{slichter:book} In Fig.~\ref{fig:ComparisonSimpleModel}, we compare $T_2$ from Eq.~(\ref{eq:Simple3x3DecRateFinalMainText}) with $T_2$ derived from Eq.~(\ref{eq:matrix}) for $\Omega = 0$ (see also Fig.~\ref{fig:different_spin_orbit}), and find excellent agreement at $T \gtrsim 50\mbox{ mK}$ where relaxation is negligible.

The above analysis provides further insight and gives explanations for the results observed in this work. First, Eq.~(\ref{eq:Simple3x3DecRateFinalMainText}) illustrates that dephasing requires two-phonon processes and cannot be achieved with a single phonon only. As dephasing leaves the energy of the electrons and the phonon bath unchanged, the single phonon would have to fulfill $\omega_{\bm{q}s} = 0 = |\bm{q}|$. However, phonons with infinite wavelengths do not affect the lifetimes, which can be explained both via $e^{i\bm{q}\cdot\bm{r}} \to 1$ [see Eq.~(\ref{eq:HelphBasicFormMainText})] and via the vanishing density of states at $\omega_{\bm{q}s} \to 0$ for acoustic phonons in bulk. Thus, $\Gamma_2^{\rm 1p} = \Gamma_1^{\rm 1p}/2$ in all our calculations, where $\Gamma_1^{\rm 1p}$ is the relaxation rate due to one-phonon processes. Second, as discussed above, we find that the hyperfine interaction in combination with electron-phonon coupling presents an important source of relaxation in this system.\cite{raith:prl12} Third, the strong dephasing at large detuning $\epsilon$ results from two-phonon processes between states of type $\ket{(1,1)S}$ and $\ket{(0,2)S}$. This mechanism is very effective because the spin state remains unchanged. Therefore, the dephasing requires neither SOI nor hyperfine coupling, and we note that Eq.~(\ref{eq:Simple3x3DecRateFinalMainText}) reveals a strong dependence of $T_\varphi$ on the tunnel coupling $t$ and the splitting $\Delta_S^\prime$. Hence, the short $T_\varphi$ in the biased DQD can be interpreted as a consequence of the Pauli exclusion principle. When the energy of the right QD is lowered ($\epsilon > 0$), the singlet state of lowest energy changes from $\ket{(1,1)S}$ toward $\ket{(0,2)S}$, since the symmetric orbital part of the wave function allows double-occupancy of the orbital ground state in the right QD. The triplet states, however, remain in the (1,1) charge configuration. While this feature allows tuning of the exchange energy and readout via spin-to-charge conversion on the one hand,\cite{petta:sci05} it enables strong dephasing via electron-phonon coupling on the other hand: effectively, phonons lead to small fluctuations in $\epsilon$; due to Pauli exclusion, these result in fluctuations of the exchange energy and, thus, in dephasing. This mechanism is highly efficient in biased DQDs, but strongly suppressed in unbiased ones, as we show in Sec.~\ref{sec:SmallDetuning} and Appendix~\ref{sec:SimpleModelDephasingZeroDetunSinglets}.

\subsection{Angular dependence}
\label{secsub:AngularDependence}

\begin{figure}[tb] 
\begin{center}
\includegraphics[width=0.85\linewidth]{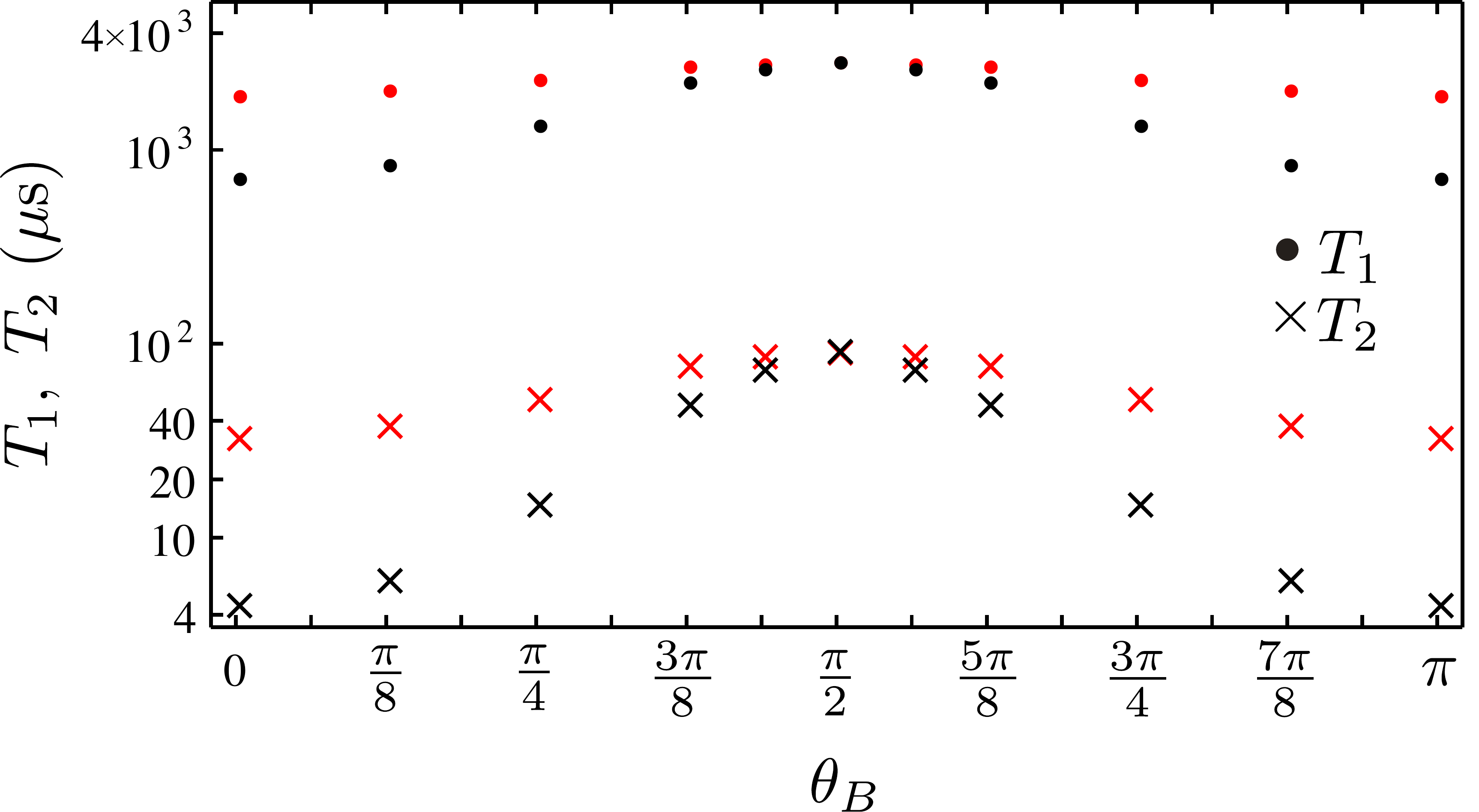}
\caption{Dependence of the relaxation $(T_1)$ and decoherence time $(T_2)$ on the angle $\theta_B$ between the in-plane magnetic field $\bm{B}$ and the $x$ axis that connects the QDs. When $\bm{B} \perp x$ ($\theta_B=\pi/2$), both $T_1$ and $T_2$ exhibit a maximum. Red (black) corresponds to the spin-orbit lengths $l_R=2\ \mu\rm m$ and $l_D=1\ \mu\rm m$ ($l_R=1\ \mu\rm m$ and $l_D=0.5\ \mu\rm m$). For the stronger SOI, the lifetimes increase by almost two orders of magnitude. For details, see text.}
\label{fig:angular_dependence}
\end{center}
\end{figure} 

We also calculate the dependence of $T_1$ and $T_2$ on the angle between $\bm{B}$ and the $x$ axis, assuming that $x \parallel [110]$. The results for $T = \mbox{100 mK}$ and $J_{\rm tot} = 1.43 \mbox{ $\mu$eV}$ are plotted in Fig.~\ref{fig:angular_dependence}. Remarkably, the phonon-induced lifetimes of the qubit are maximal when $\bm{B} \perp x$ and minimal when $\bm{B} \parallel x$. The difference between minimum and maximum increases strongly with the SOI, and for $l_R=1\mbox{ $\mu$m}$ and $l_D=0.5\mbox{ $\mu$m}$ we already expect improvements by almost two orders of magnitude. These features can be understood via the matrix elements of the effective SOI,\cite{stano:prl06, golovach:prb08, raith:prl12}
\begin{equation}
\Omega = F_{\rm SOI}(a,l_c) E_Z \frac{l_D\cos{(\theta_B - \theta)}+l_R\cos{(\theta_B + \theta)}}{l_D l_R},
\end{equation}
where $\theta_B$ ($\theta$) is the angle between $\bm{B}$ (the $x$ axis) and the crystallographic axis [110], and $F_{\rm SOI}(a,l_c)$ is a function of $a$ and $l_c$. From this result, we conclude that there always exists an optimal orientation for the in-plane magnetic field for which the effective SOI is suppressed and, thus, for which the phonon-mediated decay of the qubit state is minimal (comparing the lifetimes at fixed $J_{\rm tot}$). Remarkably, one finds for $x \parallel [110]$ ($\theta = 0$) that this suppression always occurs when $\bm{B} \perp x$ ($\theta_B = \pi/2$), independent of $l_R$ and $l_D$. In the case where $\Omega = 0$, the finite $T_2$ in our model results from admixtures with $|(0,2)S\rangle$, as explained in Sec.~\ref{secsub:OriginStrongDephasing}. Due to the hyperfine interaction, these admixtures also lead to finite $T_1$. We wish to emphasize, however, that suppression of the effective SOI only results in a substantial prolongation of the lifetimes when the spin-orbit lengths are rather short, as the dominant decay mechanism in biased DQDs is very effective even at $\Omega = 0$.

\section{Regime of Small Detuning}
\label{sec:SmallDetuning}

All previous results were calculated for a large detuning $\epsilon \sim U - V_\pm$. Now we consider an unbiased DQD, i.e., the region of very small $\epsilon$. The dominant decay mechanism in the biased DQD is strongly suppressed at $\epsilon \simeq 0$, where the basis states $\ket{(2,0)S}$ and $\ket{(0,2)S}$ are both split from $\ket{(1,1)S}$ by a large energy $U - V_{+}$. Adapting the simple model behind Eq.~(\ref{eq:Simple3x3DecRateFinalMainText}) to an unbiased DQD yields
\begin{equation}
\frac{8 t^4}{\hbar^2 (U - V_{+})^6} \int_{-\infty}^{\infty}  \langle\widetilde{P}^2(0)\widetilde{P}^2(\tau)\rangle d\tau
\end{equation}
as the associated dephasing time (see Appendix~\ref{sec:SimpleModelDephasingZeroDetunSinglets} for details). Comparing the prefactor with that of Eq.~(\ref{eq:Simple3x3DecRateFinalMainText}) results in a remarkable suppression factor below $10^{-4}$ for the parameters in this work. As explained in Appendix~\ref{sec:SimpleModelDephasingZeroDetunSinglets}, this suppression factor may also be estimated via $(\Delta_S^\prime)^4/(U-V_{+})^4$ for fixed $J_{\rm tot}$, where $\Delta_S^\prime$ is the splitting between the eigenstates of type $\ket{(1,1)S}$ and $\ket{(0,2)S}$ at large $\epsilon$ and $U-V_{+}$ is the above-mentioned splitting at $\epsilon\simeq 0$. 

\begin{figure}[tb] 
\begin{center}
\includegraphics[width=0.85\linewidth]{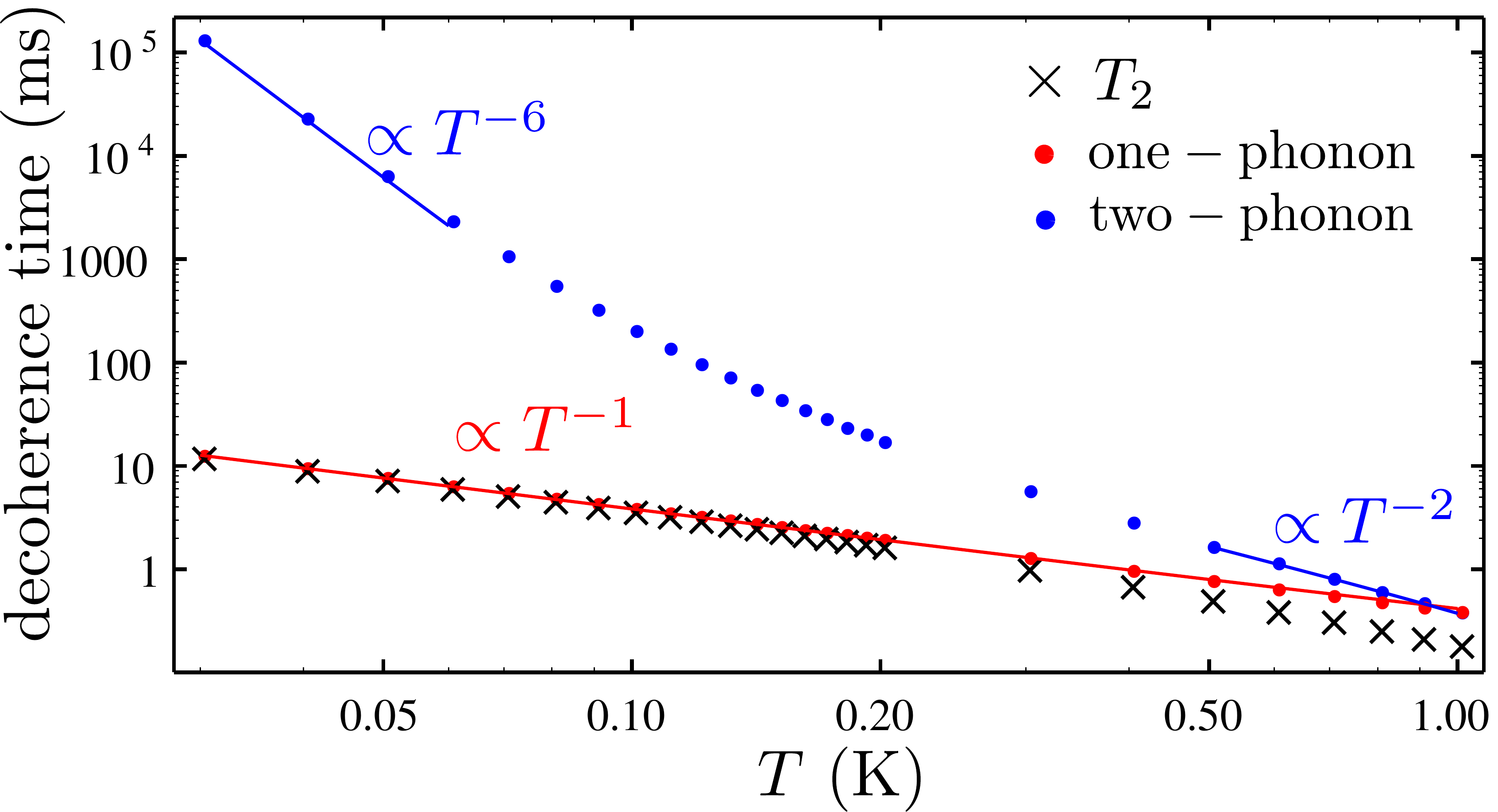}
\caption{Temperature dependence of the decoherence time $(T_2)$ and its one-phonon ($1/\Gamma_2^{\rm 1p}$) and two-phonon ($1/\Gamma_2^{\rm 2p}$) parts for the detuning $\epsilon \simeq 0$, where excited states are taken into account. For this plot $U = 1\mbox{ meV}$, $V_+ = 50\mbox{ $\mu$eV}$, $V_- = 49.5\mbox{ $\mu$eV}$, $t = 24\mbox{ $\mu$eV}$, $J_{\rm tot} = 1.41 \mbox{ $\mu$eV}$, and the other paramters as described in the text. We note that $T_2 \simeq 2 T_1$. }
\label{fig:excited_states}
\end{center}
\end{figure} 

Consequently, the lifetimes $T_1$ and $T_2$ in the unbiased DQD are no longer limited by $|(2,0)S\rangle$ or $|(0,2)S\rangle$, but by states with an excited orbital part (see Fig.~\ref{fig:spectrum}). We therefore extend the subspace by the basis states $|(1^*,1)S\rangle$, $|(1^*,1)T_0\rangle$, $|(1^*,1)T_+\rangle$, and $|(1^*,1)T_-\rangle$, and proceed analogously to the case of large detuning (see Appendixes \ref{sec:BasisStates} and \ref{sec:HamiltonianSmallDetuning} for details). The asterisk denotes that the electron is in the first excited state, leading to an energy gap of $\hbar \omega_0$ compared to the states without asterisk. Setting $\bm{B} \parallel x \parallel [110]$, the orbital excitation is taken along the $x$ axis, because states with the excitation along $y$ turn out to have negligible effects on the qubit lifetimes. From symmetry considerations, states with the excited electron in the right QD should only provide quantitative corrections of the lifetimes by factors on the order of 2 and are therefore neglected in this analysis. The resulting temperature dependence of $T_2$, $1/\Gamma_2^{\rm 1p},$ and $1/\Gamma_2^{\rm 2p}$ is shown in Fig.\ \ref{fig:excited_states}. The plotted example illustrates that two-phonon processes affect $T_2$ only at rather high temperatures when $\epsilon$ is small, leading to $T_2 \propto T^{-1}$ for a wide range of $T$ due to single-phonon processes. In stark contrast to the biased DQD, we find $T_2 \simeq 2 T_1$. Remarkably, the absolute value of $T_2$ is of the order of milliseconds, which exceeds the $T_2$ at large $\epsilon$ by 2--3 orders of magnitude. For $\bm{B} \perp x$, $x \parallel [110]$, and typical sample temperatures $T \sim 0.1\mbox{ K}$, we find that the lifetimes can be enhanced even further.

\section{Conclusions and outlook}
\label{sec:Conclusions}

In conclusion, we showed that one- and two-phonon processes can be major sources of relaxation and decoherence for $S$-$T_0$ qubits in DQDs. Our theory provides a possible explanation for the experimental data of Ref.~\onlinecite{dial:prl13}, and we predict that the phonon-induced lifetimes are prolonged by orders of magnitude at small detunings and, when the SOI is strong, at certain orientations of the magnetic field. Our results may also allow substantial prolongation of the relaxation time recently measured in resonant exchange qubits.\cite{medford:prl13}

While the model developed in this work applies to a wide range of host materials, the resulting lifetimes depend on the input parameters and, thus, on the setup and the heterostructure. By separately neglecting the deformation potential coupling ($\Xi = 0$) and the piezoelectric coupling ($h_{14 }= 0$), we find that the qubit lifetimes of Figs.~\ref{fig:tem_dep}--\ref{fig:excited_states} for GaAs DQDs are limited by the piezoelectric electron-phonon interaction, the latter providing much greater decay rates than the deformation potential coupling. Consequently, the phonon-limited lifetimes of singlet-triplet qubits may be long in group-IV materials such as Ge or Si,\cite{maune:nat12, prance:prl12, zwanenburg:rmp13} where the piezoelectric effect is absent due to bulk inversion symmetry.

Essentially, there are two different schemes for manipulating singlet-triplet qubits in DQDs electrically. The first and commonly realized approach is based on biased DQDs and uses the detuning to control the exchange energy.\cite{petta:sci05} Alternatively, the exchange energy can be controlled by tuning the tunnel barrier\cite{loss:pra98} rather than the detuning. Our results suggest that the second approach is advantageous, as it applies to unbiased DQDs for which the phonon-mediated decay of the qubit state is strongly suppressed. In addition, one finds $d J_{\rm tot}/d\epsilon \propto \epsilon$ at very small detunings $\epsilon$,\cite{burkard:prb99} which implies that not only $d J_{\rm tot}/d\epsilon \simeq 0$ but also $\langle d J_{\rm tot}/d\epsilon\rangle \simeq 0$ at $\epsilon \simeq 0$, where $\langle \cdots \rangle$ now stands for the average over some random fluctuations of $\epsilon$. Therefore, singlet-triplet qubits in unbiased DQDs are also protected against electrical noise. The latter, for instance, turned out to be a major obstacle for the implementation of high-fidelity controlled-phase gates between $S$-$T_0$ qubits.\cite{shulman:sci12} Keeping in mind that two-qubit gates for singlet-triplet qubits may also be realized with unbiased DQDs,\cite{klinovaja:prb12} we conclude that operation at $\epsilon \simeq 0$ with a tunable tunnel barrier is a promising alternative to the commonly realized schemes that require nonzero detuning. As single-qubit gates for $S$-$T_0$ qubits correspond to two-qubit gates for single-electron spin qubits, the regime $\epsilon \simeq 0$ is also beneficial for many other encoding schemes.

\begin{acknowledgments} 
We thank Peter Stano, Fabio L. Pedrocchi, Mircea Trif, James R. Wootton, Robert Zielke, Hendrik Bluhm, and Amir Yacoby for helpful discussions and acknowledge support from the Swiss NF, NCCR QSIT, S$^3$NANO, and IARPA.   
\end{acknowledgments}


\appendix

\section{Basis States}
\label{sec:BasisStates}

We consider a GaAs/AlGaAs heterostructure that contains a two-dimensional electron gas (2DEG). Electric gates on the top of the sample induce a double quantum dot (DQD) potential that confines electrons and enables the implementation of a singlet-triplet qubit. Assuming that this spin qubit is based on low-energy states of two electrons in the DQD, we consider the four states of lowest energy, 
\begin{eqnarray}
|(1,1)S\rangle &=& |\Psi_+\rangle|S\rangle, \\ 
|(1,1)T_+\rangle &=& |\Psi_-\rangle|T_+\rangle, \\
|(1,1)T_0\rangle &=& |\Psi_-\rangle|T_0\rangle, \\ 
|(1,1)T_-\rangle &=& |\Psi_-\rangle|T_-\rangle,
\end{eqnarray}
two states with a doubly occupied quantum dot (QD),
\begin{eqnarray}
|(0,2)S\rangle &=& |\Psi_R\rangle|S\rangle, \\
|(2,0)S\rangle &=& |\Psi_L\rangle|S\rangle,
\end{eqnarray}
and four additional states that feature one electron in a first excited orbital state, 
\begin{eqnarray}
|(1^*,1)S\rangle &=& |\Psi_+^{e}\rangle|S\rangle, \\ 
|(1^*,1)T_+\rangle &=& |\Psi_-^{e}\rangle|T_+\rangle, \\
|(1^*,1)T_0\rangle &=& |\Psi_-^{e}\rangle|T_0\rangle, \\ 
|(1^*,1)T_-\rangle &=& |\Psi_-^{e}\rangle|T_-\rangle,
\end{eqnarray}
as the basis in this problem.
In the notation used above, the first and second index in parentheses corresponds to the occupation number of the left and right QD, respectively. The asterisk denotes that the electron in the QD is in the first excited state. The spin part of the wave functions consists of the singlet $|S\rangle$ and the triplets $|T_0\rangle$, $|T_+\rangle$, and $|T_-\rangle$,
\begin{eqnarray}
|S\rangle &=& \frac{\ket{\uparrow\downarrow} - \ket{\downarrow\uparrow}}{\sqrt{2}}, \\
|T_0\rangle &=& \frac{\ket{\uparrow\downarrow} + \ket{\downarrow\uparrow}}{\sqrt{2}}, \\
|T_+\rangle &=& \ket{\uparrow\uparrow}, \\
|T_-\rangle &=& \ket{\downarrow\downarrow},
\end{eqnarray}
where $\uparrow$ ($\downarrow$) corresponds to an electron spin oriented along (against) the externally applied magnetic field, see Appendix~\ref{sec:Hamiltonian}. 

As the two minima in the DQD potential may be approximated by the confining potential of a 2D harmonic oscillator, the one-particle wave functions for ground and first excited states can be constructed from the eigenstates of the harmonic oscillators.\cite{burkard:prb99} Defining the growth axis of the heterostructure as the $z$ axis, we consider harmonic confinement potentials around $(x,y)=(\pm a,0)$ with $l_c=\sqrt{\hbar/(m_{\rm eff} \omega_0)}$ as the confinement length in the QDs. The $x$ axis connects the two QDs, pointing from the left to the right one. The interdot distance is $L=2a$, $m_{\rm eff}$ is the effective mass of electrons in GaAs, and $\hbar\omega_0$ is the orbital level spacing in each QD. With these definitions, the orbital parts of the 2D harmonic oscillator wave functions (ground, excited along $x$, excited along $y$) can be written as
\begin{eqnarray}
\phi_{L,R}(x,y) &=& \frac{1}{\sqrt{\pi}l_c} e^{-[(x\pm a)^2+y^2]/(2 l_c^2)}, \\
\phi_{L,R}^x(x,y) &=& \sqrt{\frac{2}{\pi l_c^4}} (x\pm a) e^{-[(x\pm a)^2+y^2]/(2 l_c^2)}, \\
\phi_{L,R}^y(x,y) &=& \sqrt{\frac{2}{\pi l_c^4}} y e^{-[(x\pm a)^2+y^2]/(2 l_c^2)}. 
\end{eqnarray}
The confining potential along the $z$ axis may be considered as a triangular potential of type
\begin{equation}
V(z) = \left\{ \begin{array}{ll}
\infty, & z < 0, \\
C z, & z > 0,
\end{array} \right.
\end{equation}
where $C$ is a positive constant with units energy/length and $z = 0$ corresponds to the interface between AlGaAs ($z<0$) and GaAs ($z>0$). The ground state in such a potential can be approximated by the Fang-Howard wave function,\cite{fang:prl66}
\begin{equation}
\phi_{\rm FH}(z) = \theta(z) \frac{z}{\sqrt{2 a_z^3}} e^{-z/(2 a_z)} ,
\label{eq:FangHoward}
\end{equation}
with $a_z$ as a positive length and 
\begin{equation}
\theta(z) = \left\{ \begin{array}{ll}
0, & z < 0, \\
1, & z > 0,
\end{array} \right.
\end{equation} 
as the Heaviside step function. The Fang-Howard wave function from Eq.\ (\ref{eq:FangHoward}) is normalized and fulfills
\begin{equation}
\bra{\phi_{\rm FH}} z \ket{\phi_{\rm FH}} = 3 a_z ,
\end{equation}
which may be interpreted as the width of the 2DEG.

Following Refs.~\onlinecite{stepanenko:prb12, burkard:prb99, stepanenko:prb03} for constructing wave functions in the DQD potential, we define overlaps between the harmonic oscillator wave functions,
\begin{eqnarray}
s &=& \langle\phi_L|\phi_R\rangle=e^{-\frac{a^2}{l_c^2}}, \\
s_x &=& \langle\phi_L^x|\phi_R^x\rangle= s \left(1-\frac{2 a^2}{l_c^2}\right), \\
s_y &=& \langle\phi_L^y|\phi_R^y\rangle = s, 
\end{eqnarray}
and
\begin{eqnarray}
g &=& \frac{1-\sqrt{1-s^2}}{s}, \\
g_x &=& \frac{1-\sqrt{1-s_x^2}}{s_x}, \\
g_y &=& \frac{1-\sqrt{1-s_y^2}}{s_y} = g.
\end{eqnarray}
Then the normalized orbital parts of the one-particle wave functions for the DQD are
\begin{eqnarray}
\Phi_{L,R}(\vec{r}) &=& \frac{\phi_{L,R}(x,y) - g\phi_{R,L}(x,y)}{\sqrt{1 - 2sg + g^2}} \phi_{\rm FH}(z) , \label{eq:PhiLR}\\
\Phi_{L,R}^{e,x}(\vec{r}) &=& \frac{\phi_{L,R}^x(x,y)-g_x\phi_{R,L}^x(x,y)}{\sqrt{1-2s_xg_x+g_x^2}} \phi_{\rm FH}(z),\\
\Phi_{L,R}^{e,y}(\vec{r}) &=& \frac{\phi_{L,R}^y(x,y) - g \phi_{R,L}^y(x,y)}{\sqrt{1 - 2s g + g^2}} \phi_{\rm FH}(z). \label{eq:PhiLRy}
\end{eqnarray} 
We note that these six states form an orthonormal set of basis states to a very good accuracy. The only nonzero scalar products among different states are $\langle\Phi_L|\Phi^{e,x}_L\rangle$, $\langle\Phi_R|\Phi^{e,x}_R\rangle$, $\langle\Phi_L|\Phi^{e,x}_R\rangle$, and $\langle\Phi_R|\Phi^{e,x}_L\rangle$. Even though there is a nonzero overlap, the absolute values of these scalar products are small ($\sim$0.01--0.1 depending on the parameters of the DQD), which indicates that Eqs.\ (\ref{eq:PhiLR}--\ref{eq:PhiLRy}) present a very good approximation for an orthonormal basis. It is, however, important to note that we set $\langle\Phi_L|\Phi^{e,x}_L\rangle$, $\langle\Phi_R|\Phi^{e,x}_R\rangle$, $\langle\Phi_L|\Phi^{e,x}_R\rangle$, and $\langle\Phi_R|\Phi^{e,x}_L\rangle$ equal to zero when calculating the matrix elements of the effective Hamiltonian later on, in order to avoid artefacts from the finite overlap of these basis states.

Given the six basis states for the orbital part of single electrons, we can construct the two-particle wave functions \cite{stepanenko:prb12, burkard:prb99}
\begin{gather}
\Psi_{\pm}(\vec{r}_1,\vec{r}_2) = \frac{\Phi_L(\vec{r}_1)\Phi_R(\vec{r}_2)\pm \Phi_R(\vec{r}_1)\Phi_L(\vec{r}_2)}{\sqrt{2}}, \\
\Psi_{\pm}^{e, \nu}(\vec{r}_1,\vec{r}_2) = \frac{\Phi_L^{e,\nu}(\vec{r}_1)\Phi_R(\vec{r}_2)\pm \Phi_R(\vec{r}_1)\Phi_L^{e,\nu}(\vec{r}_2)}{\sqrt{2}}, \hspace{0.6cm} \\
\Psi_{L,R}(\vec{r}_1,\vec{r}_2) = \Phi_{L,R}(\vec{r}_1)\Phi_{L,R}(\vec{r}_2), 
\end{gather}
where $\nu \in \{x, y\}$. The calculations for Fig.~\ref{fig:excited_states} were done with the orbital excitation along the $x$ axis only, $\Psi_{\pm}^{e} = \Psi_{\pm}^{e,x}$, because the rates resulting from $\Psi_{\pm}^{e,y}$ are much smaller than those from $\Psi_{\pm}^{e,x}$ in this setup. For some special configurations, such as $\bm{B} \parallel y$ and $x \parallel [110]$, where $\bm{B}$ is the external magnetic field, the calculations for $\Psi_{\pm}^{e} = \Psi_{\pm}^{e,y}$ lead to lifetimes similar to or even shorter than those for $\Psi_{\pm}^{e} = \Psi_{\pm}^{e,x}$, and so states with the excitation along the $y$ axis should be taken into account in these special cases. States 
of type $(1,1^*)$ with the excited electron in the right QD will change the results only by factors around 2, and therefore were not included for simplicity.

\section{Hamiltonian}
\label{sec:Hamiltonian}

The Hamiltonian of the considered system is
\begin{eqnarray}
H &=& \sum_{j=1,2} \Bigl( H_0^{(j)} + H_Z^{(j)} + H_{\rm SOI}^{(j)} + H_{\rm hyp}^{(j)} + H_{\rm el-ph}^{(j)} \Bigr) \nonumber \\ 
& &+ H_C + H_{\rm ph} , 
\label{eq:HListOfAllTerms}
\end{eqnarray}
where the index $j$ denotes the electron, $H_0$ takes into account the motion of the electron in the double dot potential, $H_Z$ is the Zeeman term, $H_{\rm SOI}$ is the spin-orbit interaction (SOI), $H_{\rm hyp}$ is the hyperfine coupling, $H_{\rm el-ph}$ is the electron-phonon interaction, $H_C$ is the Coulomb repulsion, and $H_{\rm ph}$ is the Hamiltonian of the phonon bath. Below, we discuss the contributions to $H$ in further detail.

\subsection{Hamiltonian $H_0$}

Due to $a_z \ll l_c$, the wave function along the $z$ axis is the same for all basis states in our model. The one-particle Hamiltonian $H_0$ can therefore be written as an effective 2D Hamiltonian
\begin{equation}
H_0 = \frac{p_{x}^2+p_{y}^2}{2 m_{\rm eff}} + V(x,y),
\end{equation}
where $p_{x}$ ($p_{y}$) is the momentum along the $x$ ($y$) axis and $V(x,y)$ is the confining potential in the transverse directions. The potential $V(x,y)$ is provided by the electric gates and features a finite barrier between the two QDs. It also accounts for electric fields applied along the DQD axis that effectively shift the electron energy in the left QD by the detuning $\epsilon$  compared to the right QD.

\subsection{Coulomb repulsion}

The Hamiltonian that describes the Coulomb interaction between the two electrons is
\begin{equation}
H_C=\frac{1}{4\pi \epsilon_0\epsilon_r}\frac{e^2}{|\vec{r}_1-\vec{r}_2|} ,
\end{equation}
where $e$ is the elementary positive charge, $\epsilon_0$ is the vacuum permittivity, and $\epsilon_r$ is the relative permittivity of GaAs.

\subsection{Zeeman term}

We consider an in-plane magnetic field $\vec{B}=|\vec{B}|\vec{e_B}=B\vec{e_B}$ with arbitrary orientation in the $x$-$y$ plane. Here and in the following, $\vec{e_k}$ ($\vec{e}_\eta$) stands for the unit vector along the direction of some vector $\vec{k}$ (axis $\eta$). As the 2DEG is only a few nanometers wide, orbital effects due to an in-plane magnetic field are negligible. The Hamiltonian for the Zeeman coupling reads
\begin{equation}
H_Z = \frac{E_Z}{2} \sigma_B ,
\end{equation}
where $E_Z=g \mu_B B$ is the Zeeman energy, $g$ is the in-plane $g$ factor, $\mu_B$ is the Bohr magneton, $B = |\vec{B}|$ is the magnetic field strength, and
\begin{equation}
\sigma_B = \vec{\sigma} \cdot \vec{e_B} ,
\end{equation}
with $\vec{\sigma}$ as the vector of Pauli matrices, denotes the Pauli operator for the electron spin along the magnetic field.

\subsection{Spin-orbit interaction}

We assume that the heterostructure was grown along the [001] direction, referred to as both the $z$ and $z^\prime$ direction. Consequently, the SOI due to Rashba and Dresselhaus SOI reads
\begin{equation}
H_{\rm SOI} = \alpha \left( p_{x^\prime} \sigma_{y^\prime} - p_{y^\prime} \sigma_{x^\prime} \right) + \beta \left( p_{y^\prime} \sigma_{y^\prime} - p_{x^\prime} \sigma_{x^\prime} \right)
\label{eq:SOIfor2DEG}
\end{equation}  
for a single electron, where the axes $x^\prime$ and $y^\prime$ correspond to the main crystallographic axes [100] and [010], respectively. 

Using the antihermitian operator
\begin{equation}
S_1 = i \frac{m_{\rm eff}}{\hbar}\bigl[ \alpha \left( x^\prime \sigma_{y^\prime} - y^\prime \sigma_{x^\prime} \right) + \beta \left( y^\prime \sigma_{y^\prime} - x^\prime \sigma_{x^\prime} \right) \bigr] ,
\label{eq:solutionForS1}
\end{equation} 
which fulfills the commutation relation
\begin{equation}
[S_1, H_0] = S_1 H_0 - H_0 S_1 = - H_{\rm SOI} ,
\end{equation} 
we can remove the SOI to lowest order via a unitary (Schrieffer-Wolff) transformation,\cite{khaetskii:prb00, aleiner:prl01, golovach:prl04, stano:prb05, stano:prl06, golovach:prb08, raith:prl12}
\begin{eqnarray}
\widetilde{H} &=& e^{S} H e^{-S} =  e^{\left(\sum_j S_1^{(j)}+\ldots \right)} H e^{- \left(\sum_j S_1^{(j)}+\ldots \right)}  \nonumber \\
&\simeq& \sum_{j=1,2} \Bigl( H_0^{(j)} + H_Z^{(j)} + H_{\rm hyp}^{(j)} + H_{\rm el-ph}^{(j)} \Bigr) + H_C + H_{\rm ph} \nonumber \\ 
& &+ \sum_{j=1,2} \Bigl( [S_1^{(j)}, H_Z^{(j)}] + \frac{1}{2} [S_1^{(j)}, H_{\rm SOI}^{(j)}] \Bigr) . 
\label{eq:HtildeSchriefferWolff}
\end{eqnarray}
The perturbation theory applies when both the SOI and the Zeeman coupling are weak compared to the confinement (spin-orbit length $\gg$ confinement length; Zeeman splitting $\ll$ orbital level splitting), which is well fulfilled in the system under study. Exploiting the commutation relations $[\sigma_{x^\prime}, \sigma_{y^\prime}] = 2 i \sigma_{z^\prime}$ (and analogously for cyclic permutations) of the Pauli matrices, one finds
\begin{eqnarray}
[S_1, H_Z] &=& g \mu_B \left(\bm{r}_{\rm SOI} \times \bm{B} \right) \cdot \bm{\sigma},
\label{eq:firstCorrection}
\end{eqnarray}
where we defined the SOI-dependent vector operator
\begin{equation}
\bm{r}_{\rm SOI} = \left(\frac{y^\prime}{l_R} + \frac{x^\prime}{l_D} \right) \bm{e}_{[100]} +  \left( - \frac{x^\prime}{l_R} -  \frac{y^\prime}{l_D} \right)  \bm{e}_{[010]} . 
\end{equation}
The unit vector along the $[100]$ axis, i.e., the $x^\prime$ direction, is denoted by $\bm{e}_{[100]} = \bm{e}_{x^\prime}$, and analogously for all other crystallographic directions. The spin-orbit lengths $l_R$ and $l_D$ are defined as
\begin{eqnarray}
l_R &=& \frac{\hbar}{m_{\rm eff} \alpha} , \\
l_D &=& \frac{\hbar}{m_{\rm eff} \beta} . \label{eq:lengthDresselhausSOI}
\end{eqnarray}

The contribution due to $[S_1, H_{\rm SOI}]/2$ is less important when $B$ is sufficiently large, and considering $B \sim 0.7\mbox{ T}$ \cite{bluhm:prl10, shulman:sci12} we therefore omit it in our model. Nevertheless, we provide the result for completeness,\cite{stano:prb05} 
\begin{eqnarray}
\frac{1}{2}[S_1, H_{\rm SOI}] &=& - m_{\rm eff} \left(\alpha^2 + \beta^2 \right) 
\nonumber \\
& &+ \frac{m_{\rm eff}}{\hbar} \left( \beta^2 - \alpha^2 \right) l_{z^\prime} \sigma_{z^\prime} .
\label{eq:secondCorrection}
\end{eqnarray}
Here the operator $l_{z^\prime} = \left( x^\prime p_{y^\prime}  - y^\prime p_{x^\prime}  \right)$ corresponds to the angular momentum along the axis of strong confinement. Again, orbital effects (canonical momentum $\neq$ kinetic momentum) are negligible when the magnetic field is applied in-plane.

Finally, we mention that corrections of type $[S_1, H_{\rm hyp}]$ were neglected in Eq.\ (\ref{eq:HtildeSchriefferWolff}), because $H_Z$ is assumed to be much larger than the hyperfine coupling $H_{\rm hyp}$ that we discuss next.

\subsection{Hyperfine interaction}
\label{secsub:HyperfineInteraction}

The hyperfine interaction between the electron and the nuclear spins can be described in terms of an effective magnetic field. The latter can be split into a sum field, which is present in both QDs, and a gradient field, which accounts for the difference in the hyperfine field between the dots. As the sum field is usually small compared to the external magnetic field, and, moreover, may largely be accounted for by $H_Z$, we use $H_{\rm hyp}$ to quantify the gradient field between the dots. Hence, this Hamiltonian reads 

\begin{equation}
H_{\rm hyp}=\frac{\vec{\delta b} \cdot \vec{\sigma}}{4}\left(\mathcal{P}_L - \mathcal{P}_R \right),
\end{equation}
where $\vec{\delta b}$ arises from the hyperfine field gradient between the QDs. The operators $\mathcal{P}_L$ and $\mathcal{P}_R$ are projectors for the left and right QD, respectively, and can be written as 
\begin{eqnarray}
\mathcal{P}_{L} &=& \ket{\Phi_{L}} \bra{\Phi_{L}} + \ket{\Phi_{L}^{e,x}} \bra{\Phi_{L}^{e,x}} + \ket{\Phi_{L}^{e,y}} \bra{\Phi_{L}^{e,y}} , \hspace{0.3cm} \\
\mathcal{P}_{R} &=& \ket{\Phi_{R}} \bra{\Phi_{R}} + \ket{\Phi_{R}^{e,x}} \bra{\Phi_{R}^{e,x}} + \ket{\Phi_{R}^{e,y}} \bra{\Phi_{R}^{e,y}} ,
\end{eqnarray}
for the basis states defined in Appendix \ref{sec:BasisStates}.

We note that
\begin{equation}
\bra{(1,1)S} H_{\rm hyp} \ket{(1,1)T_0} = \frac{\delta b_B}{2} ,
\end{equation}
where
\begin{equation}
\delta b_B = \vec{\delta b} \cdot \vec{e_B} 
\end{equation} 
is the component of $\vec{\delta b}$ along the external magnetic field $\bm{B}$. Because it turns out that all other matrix elements of $H_{\rm hyp}$ within the basis of Appendix \ref{sec:BasisStates} are negligible for the lifetimes of the qubit, we approximate the hyperfine coupling by
\begin{equation}
H_{\rm hyp} \simeq \frac{\delta b_B}{2} \ket{(1,1)S} \bra{(1,1)T_0} + \textrm{h.c.} ,
\end{equation}
with the hermitian conjugate abbreviated as ``h.c.''. We set $\delta b_B = - 0.14\mbox{ $\mu$eV}$ in our calculations, in good agreement with Refs.~\onlinecite{bluhm:prl10, dial:prl13}.

\subsection{Electron-phonon coupling}
\label{secsub:ElPhInteraction}

The electron-phonon interaction
\begin{equation}
H_{\rm el-ph} = H_{\rm dp} + H_{\rm pe}
\end{equation}
comprises the deformation potential coupling $H_{\rm dp}$ and the piezoelectric coupling $H_{\rm pe}$. Both mechanisms can be derived from the displacement operator, which we therefore recall first. Most of the information summarized in this appendix on electron-phonon coupling is described in great detail in Refs.~\onlinecite{cleland:book, adachi:properties, adachi:gaas, yu:book, gantmakher:book, roessler:book, benedict:prb99}, and we refer to these for further information. 

\subsubsection{Displacement operator}
\label{secsubsub:DisplacementOperator}

Acoustic phonons in an isotropic crystal (bulk) lead to the displacement operator
\begin{equation}
\bm{u} = \sum_{\bm{q}, s} \bm{e}_{\bm{q}s} \left(c_{\bm{q}s} e^{i \bm{q}\cdot\bm{r}} a_{\bm{q}s} + c^*_{\bm{q}s} e^{- i \bm{q}\cdot\bm{r}} a^\dagger_{\bm{q}s}  \right) ,
\label{eq:DisplacementOperatorGeneral}
\end{equation} 
where $c_{\bm{q}s}$ is an arbitrary coefficient with normalization condition $\left| c_{\bm{q}s} \right|^2 = \hbar / (2 \rho V \omega_{\bm{q}s})$, $\rho$ and $V$ are the density and volume of the crystal, and $\omega_{\bm{q}s}$ is the angular frequency of the acoustic phonon of type $s$ with wave vector $\bm{q}$. For the longitudinal mode $s = l$, the dispersion relation at small $q = |\bm{q}|$ is $\omega_{\bm{q}l} = q\sqrt{(\lambda + 2 \mu)/\rho} = q v_l$, while for the transverse modes $s = t_1$ and $s = t_2$ one finds $\omega_{\bm{q}t_1} = \omega_{\bm{q}t_2} = \omega_{\bm{q}t} = q\sqrt{\mu/\rho} = q v_t$, where $\lambda$ and $\mu$ are the Lam\'{e} parameters of the material and $v_l$ ($v_t$) is the speed of sound for longitudinal (transverse) waves.\cite{cleland:book} The operators $a^\dagger_{\bm{q}s}$ and $a_{\bm{q}s}$ create and annihilate a corresponding phonon, and fulfill the commutation relations $[a^\dagger_{\bm{q}s}, a^\dagger_{\bm{q}^\prime s^\prime}] = 0$, $[a_{\bm{q}s}, a_{\bm{q}^\prime s^\prime}] = 0$, and $[a_{\bm{q}s}, a^\dagger_{\bm{q}^\prime s^\prime}] = \delta_{\bm{q},\bm{q}^\prime} \delta_{s,s^\prime}$, with $\delta_{\bm{q},\bm{q}^\prime}$ and $\delta_{s,s^\prime}$ as Kronecker deltas. For each wave vector $\bm{q}$, the three real-valued polarization vectors $\bm{e}_{\bm{q}s}$ form an orthonormal basis with $\bm{e}_{\bm{q}l} \parallel \bm{q}$. The summation over $\bm{q}$ runs over all wave vectors within the first Brillouin zone. 

With a suitable choice of the polarization vectors $\bm{e}_{\bm{q}s}$, the displacement operator from Eq.\ (\ref{eq:DisplacementOperatorGeneral}) can be simplified further. We choose these vectors in such a way that the relations 
\begin{eqnarray}
\bm{e}_{-\bm{q}l} &=& -\bm{e}_{\bm{q}l} , \label{eq:DefinitionVectorRelationsStart} \\  
\bm{e}_{-\bm{q}t_1} &=& -\bm{e}_{\bm{q}t_1} , \\ 
\bm{e}_{-\bm{q}t_2} &=& \bm{e}_{\bm{q}t_2}  , \label{eq:DefinitionVectorRelationsEnd} 
\end{eqnarray}
are fulfilled. The advantages of this definition become obvious later on, when we write down the Hamiltonian for the electron-phonon coupling. In short terms, this choice allows one to define $\bm{e}_{\bm{q}l} = \bm{q}/q$ and to represent the vectors $\bm{e}_{\bm{q}s}$ via a simple right-handed basis. Setting $c_{\bm{q}s} = \sqrt{\hbar / (2 \rho V \omega_{\bm{q}s})}$, and making use of Eqs.\ (\ref{eq:DefinitionVectorRelationsStart}) to (\ref{eq:DefinitionVectorRelationsEnd}) and of the property $\omega_{-\bm{q}s} = \omega_{\bm{q}s}$, the displacement operator can be written in the convenient form
\begin{equation}
\bm{u} = \sum_{\bm{q}, s} \sqrt{\frac{\hbar}{2 \rho V \omega_{\bm{q}s}}} \bm{e}_{\bm{q}s} \left(a_{\bm{q}s} \mp_s a^\dagger_{-\bm{q}s} \right) e^{i \bm{q}\cdot\bm{r}} ,
\label{eq:DisplacementOperatorSimpler}
\end{equation}       
where
\begin{equation}
\mp_s = \left\{ \begin{array}{ll} - &\mbox{ for } s= l , t_1 ,\\ 
+ &\mbox{ for } s= t_2 .
\end{array}\right.
\end{equation}
This representation of the displacement operator, Eq.\ (\ref{eq:DisplacementOperatorSimpler}), will now be used to derive the Hamiltonian for the electron-phonon coupling. We note that the time dependence $\bm{u} \to \bm{u}(\tau)$ and $H_{\rm el-ph} \to H_{\rm el-ph}(\tau)$ in the interaction picture (see Appendix \ref{sec:BlochRedfieldTheory}) is simply obtained via $a_{\bm{q}s} \to a_{\bm{q}s}(\tau) = a_{\bm{q}s} e^{- i \omega_{\bm{q}s} \tau}$ and $a^\dagger_{\bm{q}s} \to a^\dagger_{\bm{q}s}(\tau) = a^\dagger_{\bm{q}s} e^{i \omega_{\bm{q}s} \tau}$.

It is worth mentioning how we choose the values for the speeds of sound in GaAs. The three elastic stiffness coefficients for GaAs are $c_{11} = 118$, $c_{12} = 53.5$, and $c_{44} = 59.4$, each in units of $10^9\mbox{ N/m$^2$}$. These values were taken from Ref.~\onlinecite{cleland:book} and are in very good agreement with those in, e.g., Refs.~\onlinecite{adachi:properties, ioffe:data}. It makes sense to approximate these coefficients by $\tilde{c}_{11}$, $\tilde{c}_{12}$, and $\tilde{c}_{44}$, respectively, for which the condition $\tilde{c}_{11} = \tilde{c}_{12} + 2 \tilde{c}_{44}$ of an isotropic material is fulfilled. By postulating that the relative deviation for each of the three constants should be the same, we find $\lambda = \tilde{c}_{12} = 43.5\times 10^9\mbox{ N/m$^2$}$ and $\mu = \tilde{c}_{44} = 48.3\times 10^9\mbox{ N/m$^2$}$, corresponding to a relative deviation of 18.7\%. The resulting sound velocities in the isotropic approximation are $v_l = \sqrt{\tilde{c}_{11}/\rho} = 5.1\times 10^3\mbox{ m/s}$ 
and $v_t = \sqrt{\tilde{c}_{44}/\rho} = 3.0\times 10^3\mbox{ m/s}$. We note that 
basically the same values are obtained by simply averaging over the speeds of sound along the $[100]$, $[110]$, and $[111]$ directions (longitudinal or transverse waves, respectively), as listed, for instance, in Refs.~\onlinecite{adachi:properties, ioffe:data}.
  
\subsubsection{Deformation potential coupling}

The first coupling mechanism is the deformation potential coupling. In the presence of strain, the energy of the conduction band changes. For GaAs, a cubic semiconductor with the conduction band minimum at the $\Gamma$ point, the shift of the conduction band edge is determined by the simple Hamiltonian
\begin{equation}
H_{\rm dp} = \Xi \nabla \cdot \bm{u} = \Xi (\epsilon_{xx} + \epsilon_{yy} + \epsilon_{zz}),
\label{eq:DefPotCouplingGeneral}
\end{equation}
where $\Xi$ is the hydrostatic deformation potential, $\nabla$ is the Nabla operator, and $\epsilon_{ij}$ are the strain tensor elements, which are related to the displacement via
\begin{equation}
\epsilon_{ij} = \frac{1}{2}\left(\frac{\partial u_i}{\partial x_j} + \frac{\partial u_j}{\partial x_i} \right).
\label{eq:StrainTensorElements}
\end{equation}
The trace of the strain tensor, $\nabla \cdot \bm{u} = \epsilon_{xx} + \epsilon_{yy} + \epsilon_{zz}$, corresponds to the relative change in the volume. One finds $\Xi \approx -8\mbox{ eV}$ for GaAs,\cite{adachi:gaas, vandewalle:prb89} and so compression increases the energy of the conduction band edge. Exploiting $\nabla  e^{i \bm{q}\cdot\bm{r}} = i \bm{q}  e^{i \bm{q}\cdot\bm{r}}$ and defining $\bm{e}_{\bm{q}l} = \bm{q}/q$, substitution of Eq.\ (\ref{eq:DisplacementOperatorSimpler}) into (\ref{eq:DefPotCouplingGeneral}) yields 
\begin{eqnarray}
H_{\rm dp} = i \Xi \sum_{\bm{q}} \sqrt{\frac{\hbar}{2 \rho V \omega_{\bm{q}l}}} q \left(a_{\bm{q}l} - a^\dagger_{-\bm{q}l} \right) e^{i \bm{q}\cdot\bm{r}}.
\label{eq:DefPotCouplingFinalHamiltonian}
\end{eqnarray} 
We note that only the longitudinal mode contributes to the deformation potential coupling. This is different for the piezoelectric electron-phonon interaction that we derive next. 

\subsubsection{Piezoelectric coupling}
\label{secsubsub:PiezoelectricCoupling}

In crystals without inversion symmetry, lattice vibrations (i.e., phonons) result in a finite polarization density $\bm{P}_p^{\rm phon}$ and, consequently, lead to an effective electric field $\bm{E}_p$. The latter is characterized by the equation
\begin{equation}
0 =  \epsilon_0 \bm{E}_p  + \bm{P}_p^{\rm diel}  + \bm{P}_p^{\rm phon} = \epsilon_0 \epsilon_r \bm{E}_p + \bm{P}_p^{\rm phon} ,
\label{eq:PhononElectricFieldContextNoD}
\end{equation}
where we set the electric displacement on the left-hand side to zero due to the absence of free charges in this mechanism. The vector $\bm{P}_p^{\rm diel} = \epsilon_0 (\epsilon_r - 1) \bm{E}_p$ is the polarization density induced by the field $\bm{E}_p$, $\epsilon_0$ is the vacuum permittivity, and $\epsilon_r$ is the relative permittivity of the material ($\epsilon_r \simeq 13$ in GaAs). 
In contrast to $\bm{P}_p^{\rm diel}$, the term $\bm{P}_p^{\rm phon}$ results directly from the strain that is caused by the lattice vibrations. The polarization density $\bm{P}_p^{\rm phon}$ is related to the strain tensor elements via
\begin{equation}
P_{p,i}^{\rm phon} = \sum_{j,k} h_{ijk} \epsilon_{jk} ,
\label{eq:PhononPolarization}
\end{equation}
where the $h_{ijk}$ are the elements of the third-rank piezoelectric tensor. In zinc blende structures such as GaAs, the $h_{ijk}$ take on a rather simple form,
\begin{equation}
h_{ijk} = h_{14} |\epsilon_{ijk}| = \left\{ \begin{array}{ll} h_{14} &\mbox{ for } |\epsilon_{ijk}| = 1 ,\\ 
0 &\mbox{ for } |\epsilon_{ijk}| = 0 .
\end{array}\right.
\label{eq:PiezoelectricTensor}
\end{equation}
Here $\epsilon_{ijk}$ is the Levi-Civita symbol, and the $x_i$, $x_j$, and $x_k$ related to the indices $i$, $j$, and $k$, respectively, correspond to the main crystallographic axes.

We now proceed to calculate the electric field $\bm{E}_p$ via the relation \cite{yu:book}
\begin{equation}
\bm{E}_p =  - \frac{\bm{P}_p^{\rm phon}}{\epsilon_0 \epsilon_r} ,
\label{eq:PhononElectricFieldStartCalculation}
\end{equation}
which results directly from Eq.\ (\ref{eq:PhononElectricFieldContextNoD}). In order to improve readability, we use a short-hand notation in the remainder of this subsection for convenience: $x$, $y$, and $z$ correspond to the coordinates along the main crystallographic axes, with $\bm{e}_x$, $\bm{e}_y$, and $\bm{e}_z$ as the unit vectors along the $[100]$, $[010]$, and $[001]$ directions, respectively. Substitution of Eqs.\ (\ref{eq:DisplacementOperatorSimpler}), (\ref{eq:StrainTensorElements}), (\ref{eq:PhononPolarization}), and (\ref{eq:PiezoelectricTensor}) into Eq.\ (\ref{eq:PhononElectricFieldStartCalculation}) yields
\begin{eqnarray}
\bm{E}_p &=& - \frac{i h_{14}}{\epsilon_0 \epsilon_r} \sum_{\bm{q}, s} \begin{pmatrix} q_y e^z_{\bm{q}s} + q_z e^y_{\bm{q}s} \nonumber \\ q_z e^x_{\bm{q}s} + q_x e^z_{\bm{q}s} \\ q_x e^y_{\bm{q}s} + q_y e^x_{\bm{q}s} \end{pmatrix}\\ & & \times 
\sqrt{\frac{\hbar}{2 \rho V \omega_{\bm{q}s}}} \left(a_{\bm{q}s} \mp_s a^\dagger_{-\bm{q}s} \right) e^{i \bm{q}\cdot\bm{r}} ,
\label{eq:PiezoCouplingEfull}
\end{eqnarray} 
where 
\begin{eqnarray}
\bm{q} &=& q_x \bm{e}_x + q_y \bm{e}_y + q_z \bm{e}_z , \\
\bm{e}_{\bm{q}s} &=& e^x_{\bm{q}s} \bm{e}_x + e^y_{\bm{q}s} \bm{e}_y + e^z_{\bm{q}s} \bm{e}_z ,
\end{eqnarray}
and the three components of the vector refer to the basis \{$\bm{e}_x$, $\bm{e}_y$, $\bm{e}_z$\}. The phonon-induced electric field $\bm{E}_p$ can be split into two parts, 
\begin{equation}
\bm{E}_p = \bm{E}_p^\parallel + \bm{E}_p^\perp, 
\end{equation}
where the ``longitudinal'' part
\begin{eqnarray}
\bm{E}_p^\parallel = - \frac{i h_{14}}{\epsilon_0 \epsilon_r} \sum_{\bm{q}, s} & &\frac{2 \left(q_x q_y e^z_{\bm{q}s} + q_y q_z e^x_{\bm{q}s} + q_z q_x e^y_{\bm{q}s}\right)}{q^2}  \bm{q} \nonumber \\ 
\times & &  \sqrt{\frac{\hbar}{2 \rho V \omega_{\bm{q}s}}} \left(a_{\bm{q}s} \mp_s a^\dagger_{-\bm{q}s} \right) e^{i \bm{q}\cdot\bm{r}}
\label{eq:EparallelExplicitForm}
\end{eqnarray}
contains the contributions parallel to $\bm{q}$ for each mode, while the ``transverse'' part $\bm{E}_p^\perp = \bm{E}_p - \bm{E}_p^\parallel$ comprises the remaining components perpendicular to $\bm{q}$. The longitudinal and transverse parts fulfill
\begin{eqnarray}
\nabla \times \bm{E}_p^\parallel &=& 0, \\
\nabla \cdot \bm{E}_p^\perp &=& 0,
\end{eqnarray}
respectively. As a consequence, one can write $\bm{E}_p^\parallel$ as the gradient of a scalar potential $\Phi_p$, and $\bm{E}_p^\perp$ as the curl of a vector potential $\bm{A}_p$,
\begin{eqnarray}
\bm{E}_p^\parallel &=& - \nabla \Phi_p, \label{eq:EparallelAsMinusGradPhi}\\
\bm{E}_p^\perp &=& \nabla \times \bm{A}_p.
\end{eqnarray}
From Eqs.\ (\ref{eq:EparallelExplicitForm}) and (\ref{eq:EparallelAsMinusGradPhi}), one finds
\begin{equation}
\Phi_p = \frac{h_{14}}{\epsilon_0 \epsilon_r} \sum_{\bm{q}, s} f_{\bm{q}s} \sqrt{\frac{\hbar}{2 \rho V \omega_{\bm{q}s}}} \left(a_{\bm{q}s} \mp_s a^\dagger_{-\bm{q}s} \right) e^{i \bm{q}\cdot\bm{r}}
\label{eq:PhiFormWithoutAngles}
\end{equation}
for the scalar potential, where we introduced
\begin{equation}
f_{\bm{q}s} = \frac{2 \left(q_x q_y e^z_{\bm{q}s} + q_y q_z e^x_{\bm{q}s} + q_z q_x e^y_{\bm{q}s}\right)}{q^2} .
\label{eq:fqsDefinition}
\end{equation} 
The vector potential $\bm{A}_p$ and, hence, the transverse part $\bm{E}_p^\perp$ are usually omitted for the piezoelectric electron-phonon interaction. Reasons for this omission may be inferred from Maxwell's equations. 

In accordance with common practice, we neglect the vector potential $\bm{A}_p$ in the following and consider only the scalar potential $\Phi_p$. Using an explicit representation for the unit vectors $\bm{e}_{\bm{q}s}$, the result from Eq.\ (\ref{eq:PhiFormWithoutAngles}) can be simplified further. We choose
\begin{eqnarray}
\bm{e}_{\bm{q}l} = \frac{\bm{q}}{q} &=& \begin{pmatrix} \cos\phi_{\bm{q}} \sin\theta_{\bm{q}} \\ \sin\phi_{\bm{q}} \sin\theta_{\bm{q}} \\ \cos\theta_{\bm{q}} \end{pmatrix} , \label{eq:ChoicePolarizationUnitvectorsExplicitStart} \\ 
\bm{e}_{\bm{q}t_1} &=& \begin{pmatrix} \sin\phi_{\bm{q}} \\ - \cos\phi_{\bm{q}} \\ 0 \end{pmatrix} , \\
\bm{e}_{\bm{q}t_2} &=& \begin{pmatrix} \cos\phi_{\bm{q}} \cos\theta_{\bm{q}} \\ \sin\phi_{\bm{q}} \cos\theta_{\bm{q}} \\ - \sin\theta_{\bm{q}} \end{pmatrix} , \label{eq:ChoicePolarizationUnitvectorsExplicitEnd}
\end{eqnarray}  
in agreement with Eqs.\ (\ref{eq:DefinitionVectorRelationsStart}) to (\ref{eq:DefinitionVectorRelationsEnd}), where $0\leq \phi_{\bm{q}} < 2\pi$ is the azimuthal angle and $0\leq \theta_{\bm{q}} < \pi$ is the polar angle of $\bm{q}$ in spherical coordinates. Again, the vector components in Eqs.\ (\ref{eq:ChoicePolarizationUnitvectorsExplicitStart}) to (\ref{eq:ChoicePolarizationUnitvectorsExplicitEnd}) refer to the basis \{$\bm{e}_x$, $\bm{e}_y$, $\bm{e}_z$\}, i.e., to the unit vectors for the main crystallographic directions (note the special definition of $x$, $y$, and $z$ in this subsection). Also, we note that the \{$\bm{e}_{\bm{q}l}$, $\bm{e}_{\bm{q}t_1}$, $\bm{e}_{\bm{q}t_2}$\} defined above form a right-handed, orthonormal set of basis vectors for any $\bm{q}$. With this convenient representation, which is similar to the one chosen in Ref.~\onlinecite{benedict:prb99}, the expression $f_{\bm{q}s}$ from Eq.\ (\ref{eq:fqsDefinition}) simplifies to
\begin{eqnarray}
f_{\bm{q}l} &=& 3  \cos\theta_{\bm{q}} \sin^2\theta_{\bm{q}} \sin(2\phi_{\bm{q}}) , \\
f_{\bm{q}t_1} &=&  - \sin(2 \theta_{\bm{q}})\cos(2\phi_{\bm{q}}) , \\
f_{\bm{q}t_2} &=& -\left( 3 \sin^2\theta_{\bm{q}} - 2 \right) \sin\theta_{\bm{q}} \sin(2\phi_{\bm{q}}) , 
\end{eqnarray} 
where we mention that trigonometric identities allow one to rewrite the above relations in many different ways.

Finally, the potential energy of an electron in the phonon-induced electric field, i.e., the Hamiltonian for the piezoelectric electron-phonon coupling, corresponds to
\begin{equation}
H_{\rm pe} = - e \Phi_p ,
\end{equation}
where $-e$ is the charge of the electron.

\subsection{Phonon bath}
The Hamiltonian for the phonon bath is
\begin{equation}
H_{\rm ph} = \sum_{\vec{q},s} \hbar \omega_{\vec{q}s} \left( a^\dag_{\vec{q}s} a_{\vec{q}s} + \frac{1}{2} \right),
\end{equation}
where the sum runs again over all modes $s$ and all wave vectors $\vec{q}$ within the first Brillouin zone.

\section{Model Hamiltonian at small detuning}
\label{sec:HamiltonianSmallDetuning}

As described in detail in the main text, we study the lifetimes of the singlet-triplet qubit at both small and large detuning $\epsilon$. In this appendix, we explain the details of our model at small detunings, $\epsilon \simeq 0$.

\subsection{Exchange energy and orbital level spacing}
\label{secsub:ExchangeEnergyEtc}

In the unbiased DQD, the energy of $\ket{(0,2)S}$ and $\ket{(2,0)S}$ is much larger than that of $(1^*,1)$-type states with an excited orbital part. This allows us to calculate the lifetimes with an 8$\times$8 matrix [see Eq.\ (\ref{eq:Matrix8x8zeroDetuning})] that is based on states of type $(1,1)$ and $(1^*,1)$ only. Even though $\ket{(0,2)S}$ and $\ket{(2,0)S}$ are not part of the basis, their presence can be accounted for as described below.   

Considering the basis states introduced in Appendix~\ref{sec:BasisStates} and shifting the energy globally by $\bra{(1,1)T_0} \bigl( H_0^{(1)} + H_0^{(2)} + H_C \bigr) \ket{(1,1)T_0}$, the Hamiltonian $H_0^{(1)} + H_0^{(2)} + H_C$ can be approximated via
\begin{eqnarray}
H_0^{(1)} + H_0^{(2)} + H_C &\approx& - J_S \ket{(1,1)S}\bra{(1,1)S} \hspace{1cm} \\
& &+ \Delta E \left( \ket{\Psi_{+}^{e}}\bra{\Psi_{+}^{e}} + \ket{\Psi_{-}^{e}}\bra{\Psi_{-}^{e}} \right) , \hspace{0.4cm} \nonumber
\end{eqnarray}
where the exchange energy $J_S$ results from admixtures with $\ket{(0,2)S}$ and $\ket{(2,0)S}$. The energy gap $\Delta E \simeq \hbar \omega_0$ is well described by the level spacing $\hbar \omega_0$ in the left QD and corresponds to the energy difference between the four states of lowest energy in the DQD and the states with excited orbital part.

We note that $J_S$ can be estimated \cite{stepanenko:prb12, burkard:prb99, stepanenko:prb03} by projecting $H_0^{(1)} + H_0^{(2)} + H_C$ onto the subspace $\{\ket{(2,0)S}, \ket{(0,2)S}, \ket{(1,1)S} \}$ through a projector $\mathcal{P}_{S 3}$, which yields the Hamiltonian 
\begin{equation}
H_{S3} = \mathcal{P}_{S3} \left(H_0^{(1)} + H_0^{(2)} + H_C\right) \mathcal{P}_{S3}
\end{equation}
with matrix representation 
\begin{equation}
H_{S3} = \begin{pmatrix}
U - V_- & 0 & -\sqrt{2}t \\
0 & U - V_- & -\sqrt{2} t \\ 
-\sqrt{2} t & -\sqrt{2} t & V_+ - V_-
\end{pmatrix}.
\end{equation}
Here
\begin{equation}
t = - \langle\Phi_L| H_0 |\Phi_R\rangle -\frac{1}{\sqrt{2}}\langle\Psi_+|H_C|\Psi_R\rangle 
\end{equation}
is the hopping amplitude (also referred to as the tunnel coupling), $U = \langle \Psi_R | H_C | \Psi_R \rangle$ is the on-site repulsion, $V_{\pm} = \langle \Psi_{\pm} | H_C | \Psi_{\pm}\rangle$, and the energy was globally shifted as mentioned before. Diagonalization of $H_{S3}$ results in 
\begin{eqnarray}
\widetilde{H}_{S3} &=& U_{S3}^\dagger H_{S3} U_{S3} \nonumber \\
&=& \begin{pmatrix}
U - 2 V_- + V_+ + J_S & 0 & 0 \\
0 & U - V_- & 0 \\ 
0 & 0 & - J_S
\end{pmatrix} ,
\end{eqnarray}
where $U_{S3}$ is the matrix for the unitary transformation and
\begin{equation}
J_S = \frac{1}{2}\left( \sqrt{16 t^2 + (U- V_+)^2} - U - V_+ + 2 V_- \right) 
\label{eq:Jformula}
\end{equation}
is the resulting exchange splitting between $\ket{(1,1)S}$ and $\ket{(1,1)T_0}$. Considering $\epsilon \simeq 0$, the formulas for $J_S$ and $U_{S3}$ from this estimate allow us to account for admixtures of $\ket{(2,0)S}$ and $\ket{(0,2)S}$ to the qubit state of type $\ket{(1,1)S}$ and, consequently, to study the effects of these admixtures on the phonon-induced lifetimes of the qubit.

\subsection{Matrix representation}
\label{secsub:MatrixRepresentation}

We analyze the qubit lifetimes in an unbiased DQD by projecting the Hamiltonian $\widetilde{H}$, Eq.\ (\ref{eq:HtildeSchriefferWolff}), onto the basis $\{ |(1,1)S\rangle$, $|(1,1)T_0\rangle$, $|(1,1)T_+\rangle$, $|(1,1)T_-\rangle$, $|(1^*,1)S\rangle$, $|(1^*,1)T_+\rangle$, $|(1^*,1)T_0\rangle$, $|(1^*,1)T_-\rangle \}$. The basis states are described in detail in Appendix~\ref{sec:BasisStates}, and the projection yields   
\begin{widetext}
\begin{equation}
\widetilde{H}=
\begin{pmatrix}
    -J_S + P_{SS} & \frac{\delta b_B}{2} & \frac{\Omega}{\sqrt{2}} & -\frac{\Omega}{\sqrt{2}} & P^e_{cr} & \frac{\Omega_1}{\sqrt{2}} & 0 & -\frac{\Omega_1}{\sqrt{2}} \\
    \frac{\delta b_B}{2} & P_T & 0 & 0 & 0 & -\frac{\Omega_1}{\sqrt{2}} & P^e_{cr} & -\frac{\Omega_1}{\sqrt{2}}  \\ 
    \frac{\Omega}{\sqrt{2}} & 0 & E_Z + P_T & 0 & \frac{\Omega_1}{\sqrt{2}} & P^e_{cr} & - \frac{\Omega_1}{\sqrt{2}} & 0 \\
    -\frac{\Omega}{\sqrt{2}} & 0 & 0 & -E_Z+P_T & -\frac{\Omega_1}{\sqrt{2}} & 0 & -\frac{\Omega_1}{\sqrt{2}} & P^e_{cr}\\
    P^{e\dag}_{cr} & 0 & \frac{\Omega_1}{\sqrt{2}} & -\frac{\Omega_1}{\sqrt{2}} & \Delta E +P^e & \frac{\Omega_2}{\sqrt{2}} & 0 & -\frac{\Omega_2}{\sqrt{2}}\\
    \frac{\Omega_1}{\sqrt{2}} & -\frac{\Omega_1}{\sqrt{2}} & P_{cr}^{e\dag} & 0 & \frac{\Omega_2}{\sqrt{2}} & \Delta E+E_Z+P^e & -\frac{\Omega_3}{\sqrt{2}} & 0\\
    0 & P^{e\dag}_{cr} &  - \frac{\Omega_1}{\sqrt{2}} & -\frac{\Omega_1}{\sqrt{2}} & 0 & -\frac{\Omega_3}{\sqrt{2}} & \Delta E +P^e & -\frac{\Omega_3}{\sqrt{2}} \\
    -\frac{\Omega_1}{\sqrt{2}} & -\frac{\Omega_1}{\sqrt{2}} & 0 &  P^{e\dag}_{cr} & -\frac{\Omega_2}{\sqrt{2}} & 0 & -\frac{\Omega_3}{\sqrt{2}} & \Delta E-E_Z+P^e\\
  \end{pmatrix} + H_{\rm ph} .
\label{eq:Matrix8x8zeroDetuning}
\end{equation}
\end{widetext}
Here the $\Omega$ with different indices quantify the matrix elements resulting from the SOI. Defining 
\begin{equation}
R_{\rm SOI}=(\bm{r}_{\rm SOI}\times\vec{e_B})_z ,
\end{equation}
one obtains 
\begin{eqnarray}
\Omega &=& E_Z \left( \langle \Phi_L|R_{\rm SOI}|\Phi_L\rangle-\langle \Phi_R|R_{\rm SOI}|\Phi_R\rangle \right), \\ 
\Omega_1 &=& E_Z \langle \Phi_L|R_{\rm SOI}|\Phi_L^{e,\nu}\rangle, \\ 
\Omega_2 &=& E_Z \left( \langle \Phi_L^{e,\nu}|R_{\rm SOI}|\Phi_L^{e,\nu}\rangle - \langle \Phi_R|R_{\rm SOI}|\Phi_R\rangle \right) , \hspace{0.2cm} \\ 
\Omega_3 &=& E_Z \left( \langle \Phi_L^{e,\nu}|R_{\rm SOI}|\Phi_L^{e,\nu}\rangle + \langle \Phi_R|R_{\rm SOI}|\Phi_R\rangle \right).
\end{eqnarray}
Analogously, the electron-phonon coupling is denoted by $P$ with different labels, 
\begin{eqnarray}
P_T &=& \langle\Phi_R|H_{\rm el-ph}|\Phi_R\rangle+\langle\Phi_L|H_{\rm el-ph}|\Phi_L\rangle , \\ 
P^e &=& \langle\Phi_L^{e,\nu}|H_{\rm el-ph}|\Phi_L^{e,\nu}\rangle + \langle \Phi_R|H_{\rm el-ph}|\Phi_R\rangle , \\
P_{cr}^e &=& \langle\Phi_L|H_{\rm el-ph}|\Phi_L^{e,\nu}\rangle . 
\end{eqnarray}
The above expressions for $\Omega_1$, $\Omega_2$, $\Omega_3$, $P^e$, and $P_{cr}^e$ correspond to $\Psi_{\pm}^{e} = \Psi_{\pm}^{e,\nu}$, for which the orbital excitation is chosen along the axis $\nu \in \{ x, y\}$. 

In order to account for the finite admixtures from the states $\ket{(0,2)S}$ and $\ket{(2,0)S}$, we set the matrix element $\bra{(1,1)S} \bigl(  H_{\rm el-ph}^{(1)} + H_{\rm el-ph}^{(2)} \bigr) \ket{(1,1)S}$ of the electron-phonon interaction to $P_{SS}$. The latter is a linear combination of $P_{SL}$, $P_{SR}$, $P_{S}$, and $P_{S}^\dagger$, where 
\begin{eqnarray}
P_{SL} &=& 2\langle\Phi_L|H_{\rm el-ph}|\Phi_L\rangle, \\
P_{SR} &=& 2\langle\Phi_R|H_{\rm el-ph}|\Phi_R\rangle, \\
P_{S} &=& \sqrt{2}\langle\Phi_R|H_{\rm el-ph}|\Phi_L\rangle . 
\end{eqnarray}
The coefficients of the linear combination depend on $U$, $V_+$, $V_-$, and $t$. We find these coefficients by projecting $H_{\rm el-ph}^{(1)} + H_{\rm el-ph}^{(2)}$ onto the subspace $\{\ket{(2,0)S}, \ket{(0,2)S}, \ket{(1,1)S} \}$, 
\begin{equation}
\mathcal{P}_{S3} (H_{\rm el-ph}^{(1)}+H_{\rm el-ph}^{(2)}) \mathcal{P}_{S3} = \begin{pmatrix}
P_{SL} & 0 & P_{S}^\dagger \\
0 & P_{SR} & P_{S} \\ 
P_{S} & P_{S}^\dagger & P_T
\end{pmatrix},
\end{equation}
which allows calculation of $P_{SS}$ via 
\begin{equation}
P_{SS} = \left(U_{S3}^\dagger \mathcal{P}_{S3} \bigl( H_{\rm el-ph}^{(1)}+H_{\rm el-ph}^{(2)} \bigr) \mathcal{P}_{S3} U_{S3} \right)_{33} .
\end{equation}
For further information on the transformation matrix $U_{S3}$, see Appendix~\ref{secsub:ExchangeEnergyEtc}. 

We note, however, that the above-mentioned contributions from $\ket{(2,0)S}$ and $\ket{(0,2)S}$ to $P_{SS}$ turn out to be negligibly small, because setting $P_{SS} = P_T$ does not affect the lifetimes in our calculations. Furthermore, two-phonon processes based on admixtures from $\ket{(2,0)S}$ and $\ket{(0,2)S}$ are strongly suppressed at $\epsilon \simeq 0$ and can be omitted, as we explain in detail in Appendix~\ref{sec:SimpleModelDephasingZeroDetunSinglets}. In conclusion, we find for the parameters in this work that the qubit lifetimes in unbiased DQDs are determined by the basis states with excited orbital parts. The corrections from $\ket{(2,0)S}$ and $\ket{(0,2)S}$ are negligible.

\section{Model Hamiltonian at large detuning}
\label{sec:HamiltonianLargeDetuning}

When $|\epsilon| \sim U - V_\pm$ such that the energy gap between the qubit and either $\ket{(2,0)S}$ (negative $\epsilon$) or $\ket{(0,2)S}$ (positive $\epsilon$) is smaller than the orbital level spacing, $0 < U - V_\pm - |\epsilon| < \hbar \omega_0$, the effects of higher orbitals on the lifetimes are negligible. In the regime of large detuning, we therefore project $\widetilde{H}$, Eq.\ (\ref{eq:HtildeSchriefferWolff}), onto the basis $\{\ket{(1,1)T_0}$, $\ket{(1,1)S}$, $\ket{(1,1)T_+}$, $\ket{(1,1)T_-}$, $\ket{(0,2)S}$, $\ket{(2,0)S} \}$ and investigate the lifetimes via this 6$\times$6 matrix. The explicit form of the matrix is shown in Eq.~(\ref{eq:matrix}) of the main text, and details for all its matrix elements are provided in Appendix \ref{sec:HamiltonianSmallDetuning}.

\section{Bloch-Redfield theory}
\label{sec:BlochRedfieldTheory}

Having identified a suitable matrix representation for small and large detunings, we apply a unitary transformation to $\widetilde{H}$ that diagonalizes $\widetilde{H}-\sum_{j=1,2} H_{\rm el-ph}^{(j)}$ exactly. In order to decouple the qubit subspace $\{|(1,1)S\rangle $, $|(1,1)T_0\rangle \}$ perturbatively from the remaining states, we then perform a third-order Schrieffer-Wolff transformation, leading to corrections up to the third power in the electron-phonon coupling. The perturbation theory applies when the matrix elements for the electron-phonon coupling are smaller than the energy separation between the qubit and the other states.   

The resulting effective Hamiltonian $H_{\rm eff} = H_{\rm q} + H_{\rm q-ph} + H_{\rm ph}$ for the $S$-$T_0$ qubit, its interaction with the phonon bath, and the bath itself can be described in terms of a coupled spin-1/2 system and allows application of the Bloch-Redfield theory.\cite{slichter:book, golovach:prl04, borhani:prb06} Introducing the effective magnetic fields $\vec{B_{\rm eff}}$ and $\vec{\delta B}$, we write the Hamiltonian of the qubit as
\begin{equation}
H_{\rm q} = \frac{1}{2}g \mu_B \vec{B_{\rm eff}}\cdot\vec{\sigma^\prime} ,
\end{equation}
and the Hamiltonian for the interaction between the qubit and the phonon bath reads
\begin{equation}
H_{\rm q-ph}(\tau)=\frac{1}{2}g \mu_B \vec{\delta B}(\tau)\cdot\vec{\sigma^\prime}.
\end{equation}
Here $\bm{\sigma^\prime}$ is the vector of spin-1/2 Pauli matrices for the $S$-$T_0$ qubit, $\tau$ is the time, and the time-dependent $H_{\rm q-ph}(\tau)$ is written in the interaction representation, 
\begin{equation}
H_{\rm q-ph}(\tau) = e^{i H_{\rm ph} \tau/\hbar} H_{\rm q-ph} e^{-i H_{\rm ph} \tau/\hbar} .
\end{equation} 
Next, following Refs.\ \onlinecite{golovach:prl04, borhani:prb06}, we define the spectral functions 
\begin{equation}
J_{ij}(\omega)=\frac{g^2\mu_B^2}{2\hbar^2}\int_0^\infty e^{-i\omega \tau}\langle\delta B_i(0)\delta B_j(\tau)\rangle d\tau ,
\end{equation}
where the temperature-dependent correlators $\langle \delta B_i(0)\delta B_j(\tau)\rangle$ with $i,j \in \{x,y,z\}$ are calculated for a phonon bath in thermal equilibrium. More precisely, we assume that the density matrix $\rho_{\rm ph}$ that describes the mixed state of the phonon bath is diagonal when represented via standard Fock states for the phonons considered here (i.e., occupation numbers referring to acoustic phonons classified by the wave vectors $\bm{q}$ and modes $s$), with the probabilities on the diagonal provided by Boltzmann statistics. The correlator $\langle \delta B_i(0)\delta B_j(\tau)\rangle$ corresponds to the expectation value of the operator $\delta B_i(0)\delta B_j(\tau)$ and, thus, is equal to the trace of $\rho_{\rm ph} \delta B_i(0)\delta B_j(\tau)$. In particular, one obtains $\langle a^\dagger_{\bm{q}s} a_{\bm{q}^\prime s^\prime} \rangle = \delta_{\bm{q},\bm{q}^\prime} \delta_{s,s^\prime} n_B(\omega_{\bm{q}s})$, where 
\begin{equation}
n_B(\omega) = \frac{1}{e^{\hbar \omega/(k_B T)} - 1 } 
\label{eq:BoseEinsteinDistributionAppendix}
\end{equation}
is the Bose-Einstein distribution, $k_B$ is the Boltzmann constant, and $T$ is the temperature.

Using the formulas (C16) and (C25)--(C27) from Ref.~\onlinecite{borhani:prb06}, it is possible to express the lifetimes of the qubit in terms of the above-mentioned spectral functions. For convenience, we define the basis of $\vec{\sigma^\prime}$ such that only the $z$ component of the effective magnetic field $\vec{B_{\rm eff}}$ is nonzero. In this case, the lifetimes depend solely on the quantities
\begin{eqnarray}
J^+_{ii}(\omega) &=& \mbox{Re}[J_{ii}(\omega) + J_{ii}(-\omega)] \nonumber \\
&=& \frac{g^2\mu_B^2}{2\hbar^2} \int_{-\infty}^{\infty} \cos(\omega \tau) \langle\delta B_i(0)\delta B_i(\tau)\rangle d\tau . 
\label{eq:JplusiiFormulaAppendix}
\end{eqnarray} 
The last equality holds because the $\delta B_i(\tau)$ are hermitian and the correlators are time-translational invariant. We finally calculate the relaxation time $T_1$ of the qubit via
\begin{equation}
\frac{1}{T_1} = J_{xx}^+(\omega_{Z})+J_{yy}^+(\omega_{Z}),
\end{equation}
where $\hbar \omega_Z = |g \mu_B \vec{B_{\rm eff}}|$ is the effective Zeeman splitting. The time $T_{\varphi}$ that accounts for pure dephasing is obtained through
\begin{equation}
\frac{1}{T_{\varphi}} = J_{zz}^+(0) ,
\end{equation}
and the decoherence time $T_2$ can then be expressed in terms of $T_1$ and $T_{\varphi}$, 
\begin{equation}
\frac{1}{T_2}=\frac{1}{2T_1}+\frac{1}{T_{\varphi}} .
\end{equation}

Considering one- and two-phonon processes in our calculations, the third-order contribution to $\delta B_i(0)$ [$\delta B_i(\tau)$] enters the correlator $\langle \delta B_i(0)\delta B_i(\tau)\rangle$ in Eq.~(\ref{eq:JplusiiFormulaAppendix}) together with the first-order contribution to $\delta B_i(\tau)$ [$\delta B_i(0)$]. As a consequence, the third-order terms in $\vec{\delta B}$ cannot contribute to the dephasing rate $1/T_\varphi$ (see also Appendix~\ref{sec:SimpleModelDephasing}). Furthermore, we expect only a negligible effect on the relaxation rate $1/T_1$, as the rates that arise from third-order corrections can be considered small compared to those from single-phonon processes that are based solely on the first-order terms. For simplicity, the third-order contributions to $\vec{\delta B}$ are therefore omitted in the calculations for Figs.~\ref{fig:tem_dep}--\ref{fig:excited_states}.

\section{Continuum Limit}
\label{sec:ContinuumLimit}

For the investigation of the phonon-induced lifetimes of the qubit, we consider the continuum limit and replace the summation over the phonon wave vectors $\bm{q}$ by an integral. Furthermore, the low temperatures discussed here allow integration up to infinite $q$, because the effects resulting from terms with wave vectors outside the first Brillouin zone are clearly negligible. We therefore substitute
\begin{equation}
\sum_{\bm{q}}\mbox{ } \to \mbox{ } \frac{V}{(2 \pi)^3} \int_0^\infty dq q^2 \int_0^\pi d\theta_{\bm{q}} \sin\theta_{\bm{q}} \int_0^{2 \pi} d\phi_{\bm{q}}
\end{equation} 
in our calculations. For details of the electron-phonon interaction, see Appendix~\ref{secsub:ElPhInteraction}.

\section{Simple model for dephasing at large detuning}
\label{sec:SimpleModelDephasing}

As discussed in Sec.\ \ref{secsub:OriginStrongDephasing} of the main text, the relevant dynamics at $0 < U - V_{\pm} - \epsilon < \hbar \omega_0$ and $\Omega = 0$ are very well described by the Hamiltonian
\begin{equation}
\widetilde{H} =
\begin{pmatrix}
    0 & \frac{\delta b_B}{2} & 0   \\
    \frac{\delta b_B}{2} & V_+ - V_- & - \sqrt{2}t+P_{S}^\dagger  \\ 
   0 & -\sqrt{2}t+P_{S}  & V_+ - V_- + \Delta_S + \widetilde{P}
  \end{pmatrix} + H_{\rm ph} 
\label{eq:Simple3x3StartAppendix}
\end{equation}
with basis states $\ket{(1,1)T_0}$, $\ket{(1,1)S}$, and $\ket{(0,2)S}$. Compared to Eq.~(\ref{eq:matrix}), we omitted here the decoupled states $\ket{(1,1)T_+}$, $\ket{(1,1)T_-}$, and $\ket{(2,0)S}$, subtracted $P_T$ from the diagonal (global shift, no effect on the lifetimes), and introduced
\begin{equation}
\widetilde{P} = P_{SR} - P_T
\end{equation}
as a matrix element for the electron-phonon coupling and
\begin{equation}
\Delta_S = U - V_+ - \epsilon
\end{equation}   
as the bare splitting between $\ket{(1,1)S}$ and $\ket{(0,2)S}$.

The hyperfine coupling, $\delta_B$, is the only mechanism in Eq.~(\ref{eq:Simple3x3StartAppendix}) that couples the spin states and, hence, is crucial for the relaxation of the $S$-$T_0$ qubit. In fact, we find for the parameters in this work that the relaxation times $T_1$ are mainly determined by the hyperfine coupling rather than the SOI. In order to derive a simple model for the short decoherence times [$T_2 \ll T_1$, Fig.\ \ref{fig:tem_dep}(a)], we neglect $\delta_B$ in the following, resulting in pure dephasing, and so $T_2 = T_\varphi$. Furthermore, we find that the matrix element $P_S$ is negligible for our parameter range. Defining
\begin{equation}
\widetilde{H} = H_{\rm s} + H_{\rm s-ph} + H_{\rm ph} 
\end{equation}
and omitting $\delta_B$ and $P_S$, one obtains
\begin{equation}
H_{\rm s} =
\begin{pmatrix}
    0 & 0 & 0   \\
    0 & V_+ - V_- & - \sqrt{2}t \\ 
   0 & -\sqrt{2}t & V_+ - V_- + \Delta_S 
\end{pmatrix} 
\end{equation}
for the part that describes the electronic system, and
\begin{equation}
H_{\rm s-ph} =
\begin{pmatrix}
    0 & 0 & 0   \\
    0 & 0 & 0 \\ 
   0 & 0 & \widetilde{P} 
\end{pmatrix} 
\end{equation}
for the interaction with the phonon bath.  

The Hamiltonians $H_{\rm s}$ and $H_{\rm s-ph}$ can be rewritten in a different basis $\{\ket{(1,1)T_0}$, $\ket{(1,1)S^\prime}$, $\ket{(0,2)S^\prime}\}$ as 
\begin{equation}
H_{\rm s} = 
\begin{pmatrix}
    0 & 0 & 0   \\
    0 & - J_{\rm tot} & 0 \\ 
   0 & 0 & - J_{\rm tot} + \Delta_S^\prime 
\end{pmatrix} 
\end{equation}
and
\begin{equation}
H_{\rm s-ph} = \widetilde{P}
\begin{pmatrix}
    0 & 0 & 0   \\
    0 & v_{s^\prime d}^2 & v_{s^\prime d} v_{d^\prime d} \\ 
   0 & v_{s^\prime d} v_{d^\prime d} & v_{d^\prime d}^2 
\end{pmatrix} ,
\label{eq:Simple3x3SysPhononAfterTransf}
\end{equation}
where
\begin{equation}
\Delta_S^\prime = \sqrt{\Delta_S^2 + 8 t^2}
\end{equation}
and
\begin{equation}
J_{\rm tot} = V_{-} - V_{+} +\frac{\Delta_S^\prime - \Delta_S}{2} .
\end{equation}
The basis states
\begin{eqnarray}
\ket{(1,1)S^\prime} &=& v_{s^\prime s} \ket{(1,1)S} + v_{s^\prime d} \ket{(0,2)S} , \\
\ket{(0,2)S^\prime} &=& v_{d^\prime s} \ket{(1,1)S} + v_{d^\prime d} \ket{(0,2)S}, 
\end{eqnarray}
are normalized eigenstates of $H_{\rm s}$. The notation $\ket{(1,1)S^\prime}$ and $\ket{(0,2)S^\prime}$ is justified because we consider $\Delta_S > 0$, and so $|v_{s^\prime s}|^2 > 1/2$ and $|v_{d^\prime d}|^2 > 1/2$. In Eq.~(\ref{eq:Simple3x3SysPhononAfterTransf}), $v_{s^\prime d}$ and $v_{d^\prime d}$ are assumed to be real. A suitable choice for the coefficients is, e.g.,
\begin{eqnarray}
v_{s^\prime s} &=& \frac{\Delta_S + \Delta_S^\prime}{D_{+}} , \\
v_{s^\prime d} &=& \frac{2 \sqrt{2} t}{D_{+}} , \label{eq:Simple3x3AppVsprd} \\
v_{d^\prime s} &=& \frac{\Delta_S - \Delta_S^\prime}{D_{-}} , \\
v_{d^\prime d} &=& \frac{2 \sqrt{2} t}{D_{-}} \label{eq:Simple3x3AppVdprd} , 
\end{eqnarray}
where the denominator
\begin{equation}
D_{\pm} = \sqrt{\left( \Delta_S \pm \Delta_S^\prime \right)^2 + 8t^2}
\label{eq:Simple3x3AppDenominator}
\end{equation}
ensures normalization.  

Following the steps explained in Appendix~\ref{sec:BlochRedfieldTheory}, one finds 
\begin{equation}
g \mu_B B_{\textrm{eff},z} = J_{\rm tot}
\end{equation}
and
\begin{eqnarray}
g \mu_B \delta B_{z} &=& - v_{s^\prime d}^2 \widetilde{P} + \frac{v_{s^\prime d}^2 v_{d^\prime d}^2}{\Delta_S^\prime}\widetilde{P}^2 \nonumber \\
& &+ \frac{v_{s^\prime d}^2 v_{d^\prime d}^2 \left(v_{s^\prime d}^2 - v_{d^\prime d}^2 \right)}{(\Delta_S^{\prime})^2}\widetilde{P}^3
\label{eq:Simple3x3DeltaBzAppendix}
\end{eqnarray} 
from the third-order Schrieffer-Wolff transformation. We recall that $\delta B_{x} = 0 = \delta B_{y}$ due to omission of the hyperfine coupling, and so $T_2 = T_\varphi$ (pure dephasing). Furthermore, we note that the Bloch-Redfield theory requires $\langle \bm{\delta B}(\tau) \rangle$ to vanish.\cite{slichter:book} Therefore, terms of type $a_{\bm{q}s}^\dagger a_{\bm{q}s}$ and $a_{\bm{q}s} a_{\bm{q}s}^\dagger$ must be removed from the second-order contributions to $\bm{\delta B}$ and, consequently, from the part $\propto \widetilde{P}^2$ in Eq.~(\ref{eq:Simple3x3DeltaBzAppendix}). The terms removed from $\bm{\delta B}$ can be considered as minor corrections to $\bm{B_{\rm eff}}$, with $a_{\bm{q}s}^\dagger a_{\bm{q}s} \to n_B(\omega_{\bm{q}s})$ and $a_{\bm{q}s} a_{\bm{q}s}^\dagger \to n_B(\omega_{\bm{q}s}) + 1$, where $n_B(\omega)$ is the Bose-Einstein distribution, Eq.~(\ref{eq:BoseEinsteinDistributionAppendix}). In this work, we simply neglect these corrections to $\bm{B_{\rm eff}}$ because of their smallness.

The decoherence time $T_2 = T_\varphi$ is calculated via
\begin{equation}
\frac{1}{T_2} = \frac{g^2\mu_B^2}{2\hbar^2} \int_{-\infty}^{\infty}  \langle\delta B_z(0)\delta B_z(\tau)\rangle d\tau , 
\label{eq:Simple3x3DecRateGeneralAppendix}
\end{equation}
see Appendix~\ref{sec:BlochRedfieldTheory}. Remarkably, the only nonzero contribution after insertion of Eq.~(\ref{eq:Simple3x3DeltaBzAppendix}) into Eq.~(\ref{eq:Simple3x3DecRateGeneralAppendix}) is
\begin{equation}
\frac{1}{T_2} = \frac{v_{s^\prime d}^4 v_{d^\prime d}^4}{2\hbar^2 (\Delta_S^\prime)^2} \int_{-\infty}^{\infty}  \langle\widetilde{P}^2(0)\widetilde{P}^2(\tau)\rangle d\tau . 
\label{eq:Simple3x3DecRateReducedAppendix}
\end{equation}
In particular, one finds that single-phonon processes cannot lead to dephasing,
\begin{equation}
\int_{-\infty}^{\infty} \langle\widetilde{P}(0)\widetilde{P}(\tau)\rangle d\tau = 0 .
\label{eq:Simple3x3NoSinglePhonProcAppendix}
\end{equation}
As there is no energy transfer between the electrons and the phonon bath (evaluation of $J_{zz}^{+}(\omega)$ at $\omega=0$), the left-hand side of Eq.~(\ref{eq:Simple3x3NoSinglePhonProcAppendix}) can only be nonzero for a phonon with $\omega_{\bm{q}s} = 0 = q$, for which, however, the expression vanishes as well. An analogous explanation applies to
\begin{equation}
\int_{-\infty}^{\infty} \langle\widetilde{P}^3(0)\widetilde{P}(\tau)\rangle d\tau
= 0 =
\int_{-\infty}^{\infty} \langle\widetilde{P}(0)\widetilde{P}^3(\tau)\rangle d\tau.
\end{equation} 
Consequently, the dephasing in our model results purely from two-phonon processes that are based on the second-order contributions to $\delta B_z$.

Finally, using Eqs.~(\ref{eq:Simple3x3AppVsprd}) and (\ref{eq:Simple3x3AppVdprd}) in Eq.~(\ref{eq:Simple3x3DecRateReducedAppendix}) yields
\begin{equation}
\frac{1}{T_2} = \frac{2 t^4}{\hbar^2 (\Delta_S^\prime)^6} \int_{-\infty}^{\infty}  \langle\widetilde{P}^2(0)\widetilde{P}^2(\tau)\rangle d\tau . 
\label{eq:Simple3x3DecRateFinalAppendix}
\end{equation}
We note that in the case of $|t| \ll \Delta_S$ and negligibly small $V_+ - V_-$, one finds $J_{\rm tot} \simeq 2 t^2/\Delta_S^\prime$ in this model and
\begin{equation}
\frac{2 t^4}{\hbar^2 (\Delta_S^\prime)^6} \simeq \frac{J_{\rm tot}^2}{2 \hbar^2 (\Delta_S^\prime)^4}
\label{eq:Simple3x3PrefactorEstimateLargeDet}
\end{equation} 
for the prefactor.

\section{Dephasing via singlet states at small detuning}
\label{sec:SimpleModelDephasingZeroDetunSinglets}

In order to estimate the dephasing due to the states $\ket{(2,0)S}$ and $\ket{(0,2)S}$ in an unbiased DQD, $\epsilon \simeq 0$, we study a model similar to that of Appendix~\ref{sec:SimpleModelDephasing}. Using $\ket{(2,0)S}$, $\ket{(0,2)S}$, $\ket{(1,1)S}$, and $\ket{(1,1)T_0}$ as the basis states, we consider
\begin{equation}
H_{\rm s} =
\begin{pmatrix}
    U - V_{-} & 0 & - \sqrt{2} t & 0   \\
    0 & U - V_{-} & - \sqrt{2} t & 0  \\ 
   - \sqrt{2}t & - \sqrt{2}t & V_+ - V_- & 0 \\
    0 & 0 & 0 & 0
  \end{pmatrix}
\label{eq:Simple4x4DephasingZeroDetHs}
\end{equation}
as the Hamiltonian for the electronic system and
\begin{equation}
H_{\rm s-ph} =
\begin{pmatrix}
    - \widetilde{P} & 0 & 0 & 0   \\
    0 & \widetilde{P} & 0 & 0  \\ 
    0 & 0 & 0 & 0 \\
    0 & 0 & 0 & 0
  \end{pmatrix}
\label{eq:Simple4x4DephasingZeroDetHsph}
\end{equation}
as the electron-phonon interaction. Again, we removed here $P_T$ from the diagonal and neglected the off-diagonal matrix elements $P_S$ and $P_S^\dagger$. Furthermore, we exploited the relation
\begin{equation}
P_{SL} - P_{T} = - (P_{SR} - P_{T}) = - \widetilde{P}.
\end{equation}
This relation is based on the properties
\begin{eqnarray}
\bra{\Phi_L} \cos(\bm{q}\cdot\bm{r}) \ket{\Phi_L} &=& \bra{\Phi_R} \cos(\bm{q}\cdot\bm{r}) \ket{\Phi_R} , \\
\bra{\Phi_L} \sin(\bm{q}\cdot\bm{r}) \ket{\Phi_L} &=& - \bra{\Phi_R} \sin(\bm{q}\cdot\bm{r}) \ket{\Phi_R} .
\end{eqnarray}  
Using the states $\ket{\Phi_{L,R}}$ defined in Appendix~\ref{sec:BasisStates}, Eq.~(\ref{eq:PhiLR}), it is straightforward to show that these equations apply to our calculations (at least in very good approximation, given the small width of the 2DEG). Proceeding analogously to Appendix~\ref{sec:SimpleModelDephasing} and exploiting $|t| \ll U - V_{+}$, the calculation of $T_2 = T_\varphi$ with Eqs.~(\ref{eq:Simple4x4DephasingZeroDetHs}) and (\ref{eq:Simple4x4DephasingZeroDetHsph}) yields 
\begin{equation}
\frac{1}{T_2} = \frac{8 t^4}{\hbar^2 (U - V_{+})^6} \int_{-\infty}^{\infty}  \langle\widetilde{P}^2(0)\widetilde{P}^2(\tau)\rangle d\tau ,
\label{eq:Simple4x4DecRateFinalZeroDetAppendix} 
\end{equation}
which is formally equivalent to Eq.~(\ref{eq:Simple3x3DecRateFinalAppendix}).

Operation of the qubit at $\epsilon \simeq 0$ requires control over the tunnel coupling $t$, which can be achieved by changing the tunnel barrier of the DQD with electric gates.\cite{loss:pra98} Consequently, the value of $t$ at $\epsilon \simeq 0$ is usually different from that at large $\epsilon$. As a simple estimate, using $|t| \ll U - V_{+}$ and assuming that $V_+ - V_-$ and $\delta_B$ are negligible, one finds $J_{\rm tot} \simeq 4 t^2/(U - V_{+})$ through Taylor expansion of $J_S$, Eq.~(\ref{eq:Jformula}). Analogously, one obtains
\begin{equation}
\frac{8 t^4}{\hbar^2 (U - V_{+})^6} \simeq \frac{J_{\rm tot}^2}{2 \hbar^2 (U - V_{+})^4}
\end{equation}
for the prefactor in Eq.~(\ref{eq:Simple4x4DecRateFinalZeroDetAppendix}). Considering $J_{\rm tot}$ to be the same in the biased and unbiased DQD, comparison with Eq.~(\ref{eq:Simple3x3PrefactorEstimateLargeDet}) yields a suppression factor on the order of $(\Delta_S^\prime)^4/(U - V_{+})^4$. For the parameters in this work, the associated dephasing times at $\epsilon \simeq 0$ are therefore several orders of magnitude longer than those at large $\epsilon$. The strong suppression allows omission of this mechanism in our model for an unbiased DQD described in Appendix~\ref{sec:HamiltonianSmallDetuning}. 

The matrix elements $P_S$ and $P_S^\dagger$ of the electron-phonon interaction provide a direct coupling between the state $\ket{(1,1)S}$ and the states $\ket{(0,2)S}$ and $\ket{(2,0)S}$. Consequently, these matrix elements enable dephasing via two-phonon processes even at $t = 0$. In the case of large detuning $\epsilon$, the effect of $P_S$ and $P_S^\dag$ on the dephasing time $T_\varphi$ (and on the lifetimes in general) turns out to be negligible. At $\epsilon \simeq 0$, this two-phonon-based contribution to $T_\varphi$ is suppressed even further, by a factor on the order of $4 \Delta_S^2/(U - V_{+})^2$, and can therefore be neglected in the calculation with excited orbital states (see Appendix~\ref{sec:HamiltonianSmallDetuning}).

\section{Summary of input parameters}
\label{sec:InputParameters}

Table \ref{tab:SummaryParameters} lists the values that were used for the results discussed in the main text. We note that the results are independent of the sample volume $V$, because the volume cancels out in the calculation.

\begin{table}[tp]
\caption{Input parameters used for the calculations in the main text.}
\begin{tabular}{|c|c|c|}
\hline \hline
{\bf Parameter} & {\bf Value} & {\bf References} \\ \hline 
\mbox{ }\vspace{-0.31cm} & &  \\
$\epsilon_r$ & $13$ &  \\
$\rho$ & $5.32\mbox{ g/cm$^3$}$ &  \\
$v_l$ & $5.1 \times 10^3 \mbox{ m/s}$ & \onlinecite{adachi:properties, ioffe:data, cleland:book}, Appendix \ref{secsubsub:DisplacementOperator}   \\
$v_t$ & $3.0 \times 10^3 \mbox{ m/s}$ & \onlinecite{adachi:properties, ioffe:data, cleland:book}, Appendix \ref{secsubsub:DisplacementOperator}   \\
$\Xi$ & $-8\mbox{ eV} $ & \onlinecite{adachi:gaas, vandewalle:prb89}  \\
$h_{14}$ & $-0.16\mbox{ As/m$^2$}$ & \onlinecite{adachi:properties, huebner:pss73, ioffe:data} \\
$g$ & $-0.4 $ & \\
$B$ & $0.7\mbox{ T}$ & \onlinecite{bluhm:prl10, shulman:sci12} \\
$m_{\rm eff}$ & $6.1\times 10^{-32}\mbox{ kg} $ &  \\
$\Delta E = \hbar \omega_0$ & $ 124 \mbox{ $\mu$eV} $ & \onlinecite{dial:prl13} \\ 
$l_D$ & $1,\ 0.8,\ 0.5 \mbox{ $\mu$m}$ & \onlinecite{hanson:rmp07, khaetskii:prb01, winkler:book}, Appendix \ref{sec:InputParameters} \\
$l_R$ & $2,\ 1.6,\ 1 \mbox{ $\mu$m}$ & \\
$3 a_z$ & $6\mbox{ nm} $ & Appendix \ref{sec:InputParameters} \\
$L=2a$   & $400\mbox{ nm}$  &  \\  
$\delta b_B$   & $-0.14\mbox{ $\mu$eV}$  & \onlinecite{bluhm:prl10, dial:prl13} \\  
$U$ & $ 1 \mbox{ meV} $ & \onlinecite{stepanenko:prb12} \\ 
$V_+$ & $ 40,\ 50 \mbox{ $\mu$eV} $ & \onlinecite{stepanenko:prb12} \\
$V_-$ & $ 39.78,\ 49.5 \mbox{ $\mu$eV} $ & \onlinecite{stepanenko:prb12}, Appendix \ref{sec:InputParameters} \\
$t$ & $ 7.25,\ 24 \mbox{ $\mu$eV} $ & \onlinecite{stepanenko:prb12}, Appendix \ref{sec:InputParameters} \\
\hline \hline
\end{tabular}
\label{tab:SummaryParameters}
\end{table} 

It is worth mentioning that the values $l_D \sim \mbox{0.5--1 $\mu$m}$ \cite{hanson:rmp07, khaetskii:prb01, winkler:book} for the Dresselhaus SOI are consistent with the assumed width of the 2DEG. Neglecting orbital effects, the general form of the Dresselhaus SOI for an electron in GaAs is
\begin{equation}
H_D = b_{41}^{\rm 6c6c} \left[\left(k_{y'}^2 - k_{z'}^2 \right) k_{x'} \sigma_{x'} + \mbox{ c.p.}\right] ,
\end{equation} 
where $\hbar k_{i}$ is the momentum along the $i$ axis, $\sigma_{i}$ is the corresponding Pauli operator for spin 1/2, the axes $x'$, $y'$, and $z'$ are the main crystallographic axes [100], [010], and [001], respectively, ``c.p.'' stands for cyclic permutations, and $b_{41}^{\rm 6c6c} \simeq 28 \mbox{ \AA$^3$eV}$.\cite{winkler:book} For our 2DEG with strong confinement along the [001] direction ($z$ axis), the Dresselhaus SOI can be well approximated by 
\begin{equation}
H_D \simeq b_{41}^{\rm 6c6c} \bra{\phi_{\rm FH}} k_z^2 \ket{\phi_{\rm FH}} \left( k_{y'} \sigma_{y'} - k_{x'} \sigma_{x'} \right) ,
\end{equation}
where $z' = z$ and $\phi_{\rm FH}(z)$ is the Fang-Howard wave function of Eq.\ (\ref{eq:FangHoward}). Using $\bra{\phi_{\rm FH}} k_z^2 \ket{\phi_{\rm FH}} = 1/(4 a_z^2)$, one finds
\begin{equation}
l_D \simeq \frac{4 \hbar^2 a_z^2}{m_{\rm eff} b_{41}^{\rm 6c6c}} 
\label{eq:lengthDresselhausSOIforEstimate}
\end{equation} 
from comparison with Eqs.\ (\ref{eq:SOIfor2DEG}) and (\ref{eq:lengthDresselhausSOI}). With $m_{\rm eff} = 0.067 m_{\rm el}$ \cite{winkler:book} as the effective electron mass in GaAs and $m_{\rm el}$ as the bare electron mass, evaluation of Eq.\ (\ref{eq:lengthDresselhausSOIforEstimate}) with $3 a_z = 6\mbox{ nm}$ yields $l_D \simeq 0.65\mbox{ $\mu$m}$, in good agreement with the values used in the calculation.  

The splitting between the eigenstates of type $\ket{(1,1)S}$ and $\ket{(1,1)T_0}$ after diagonalization is denoted by $J_{\rm tot} = \hbar \omega_Z$. When $J_{\rm tot} \gg |\delta b_B|$, the spin states of these eigenstates are $\ket{S}$ and $\ket{T_0}$ with high accuracy, and the state of the $S$-$T_0$ qubit precesses around the $z$ axis of the Bloch sphere. When the splitting is provided by the hyperfine coupling $\delta b_B$ instead of the exchange interaction, the eigenstates are of type $\ket{\uparrow \downarrow}$ and $\ket{\downarrow \uparrow}$, leading to precessions around the $x$ axis. In experiments, $J_{\rm tot} \gg |\delta b_B|$ is commonly realized for a biased DQD (large detuning) and the hyperfine coupling dominates in the unbiased case.\cite{petta:sci05, dial:prl13} In order to account for this feature, we set the parameters in Sec.~\ref{sec:LargeDetuning} such that $J_{\rm tot}$ at $\epsilon \simeq 0$ would be largely provided by $\delta b_B$. Using $U$, $V_+$, $V_{-}$, and $t$ approximately as in Ref.\ \onlinecite{stepanenko:prb12}, we do this by adapting $t$ (or $V_-$) such that $J_S \ll |\delta b_B|$, where $J_S$ is the bare exchange splitting at $\epsilon = 0$, Eq.\ (\ref{eq:Jformula}). The lifetimes in Figs.~\ref{fig:tem_dep}--\ref{fig:angular_dependence} were calculated with $U = 1\mbox{ meV}$, $V_+ = 40 \mbox{ $\mu$eV}$, $V_- = 39.78 \mbox{ $\mu$eV}$, and $t = 7.25 \mbox{ $\mu$eV}$, for which $J_S \ll |\delta b_B|$ is fulfilled. The detuning $\epsilon \sim 0.9\mbox{ meV}$ in these calculations was chosen such that $J_{\rm tot} = 1.43\mbox{ $\mu$eV}$, and we note that the excited states are negligible due to $0< U - V_{\pm} - \epsilon < \hbar \omega_0$. In Fig.~\ref{fig:excited_states}, where we consider operation at small detuning, the parameters $U = 1\mbox{ meV}$, $V_+ = 50 \mbox{ $\mu$eV}$, and $V_- = 49.5 \mbox{ $\mu$eV}$ are similar to before. However, in order to achieve $J_{\rm tot} = 1.41\mbox{ $\mu$eV}$ at $\epsilon \simeq 0$, we use a larger tunnel coupling $t = 24 \mbox{ $\mu$eV}$. Experimentally, this can be realized by tuning the tunnel barrier of the DQD electrically.\cite{loss:pra98}

\end{document}